\documentclass[onecolumn]{article}

\usepackage{graphicx,color}
\graphicspath{{figures-pdf/}}
\usepackage{algorithm}
\usepackage{algpseudocode}
\usepackage{multirow,bigstrut}
\usepackage{tabularx}
\usepackage{amssymb,amsmath,bm}
\usepackage{mathptmx}
\usepackage{cite}
\usepackage{url}
\usepackage{amssymb,amsmath,amsthm,bm}
\usepackage[retainorgcmds]{IEEEtrantools}

\newlength\figwidth
\newlength\imagewidth

\newlength\figsep
\usepackage{hyperref}
\hypersetup{
 linktocpage=true,
 pdfborderstyle={/S/S/W 1},
 hyperindex=true,
 bookmarks=true,
 bookmarksopen=true,
 bookmarksnumbered=true,
}

\usepackage{lineno}

\usepackage[a4paper,left=3cm,right=3cm,top=3cm,bottom=3cm]{geometry}

\usepackage[normalem]{ulem}

\begin{document}

\title{Design and ARM-embedded implementation of a chaotic map-based multicast scheme for multiuser speech wireless communication%
}


\author{\small Qiuye Gan\textsuperscript{1}, Simin Yu\textsuperscript{1}, Chengqing Li\textsuperscript{2},
Jinhu L\"u\textsuperscript{3}, Zhuosheng Lin\textsuperscript{1}, and Ping Chen\textsuperscript{1}}

\date{\small
\textsuperscript{1}College of Automation, Guangdong University of Technology, Guangzhou, China (siminyu@163.com)\\
\textsuperscript{2}College of Information Engineering, Xiangtan University, Xiangtan 411105, Hunan, China (DrChengqingLi@gmail.com)\\
\textsuperscript{3} Academy of Mathematics and Systems Sciences, Chinese Academy of Sciences, Beijing 100190, China (jhlu@iss.ac.cn)}

\maketitle

\begin{abstract}
This paper proposes a chaotic map-based multicast scheme for multiuser speech wireless communication and implements it in an ARM platform. The scheme compresses the digital audio signal decoded by a sound card and then encrypts it with a three-level chaotic encryption scheme. First, the position of every bit of the compressed data is permuted randomly with a pseudo-random number sequence (PRNS) generated by a 6-D chaotic map. Then, the obtained data are further permuted in the level of byte with a PRNS generated by a 7-D chaotic map. Finally, it is operated with a multiround chaotic stream cipher. The whole system owns the following merits: the redundancy in the original audio file is reduced effectively and the corresponding unicity distance is increased; the balancing point between a high security level of the system and real-time conduction speed is achieved well. In the ARM implementation, the framework of communication of multicast¨Cmultiuser in a subnet and the Internet Group Manage Protocol is adopted to obtain the function of communication between one client and other ones. Comprehensive test results were provided to show the feasibility and security performance of the whole system.

\textbf{keywords and phrases:} ARM-embedded implementation, multicast-multiuser, chaotic map, secure communication, speech, WIFI.
\end{abstract}

\section{Introduction}

With popularization of speech (voice and audio) record devices and development of network transmission
techniques, the security of speech transmission is becoming more and more important. To cope
with the challenge, a great number of encryption schemes based on chaos were proposed to protect
speech data. It is well known that some significant properties of chaos, such as positive Lyapunov exponents,
ergodicity, quasi-randomness, sensitive dependence on initial conditions, and system parameters,
have suggested chaotic dynamics being a promising alternative for conventional cryptographic algorithm \cite{YuLuChen13,Baptista13,LiuZhuLu15}.
The function of chaos used in encryption can be categorized as the following classes: 1) generation of scrambling (permutation) relation \cite{Licq:hierarchical:SP2016,Cqli:Fridrich:SP2017}; 2) producing pseudo-random number sequence (PRNS) \cite{Li:hyperchaotic:ND2013,tongxj:2015}; 3) realization of secure communication via chaos synchronization \cite{Celikovsky:synchronize:TAC2005,Sheu:fractional:ND2011,LiuWuLuZhu16}; 4) construction of public encryption scheme \cite{Bose:public:PRL05}. Meanwhile, some cryptanalytic works pointed out that some speech encryption schemes are insecure against the corresponding specific attack methods \cite{Li:BSS:TCASI2008,ZhuLuV12}. To counteract dynamical degradation of digital chaotic systems, various methods were proposed to weaken or even eliminate the negative influence caused by the digitization \cite{YuTangLuChen10,ShenYuLuChen15,WangYuLiLu16}. Some methods based on complex networks were proposed to measure the real degradation degree of the counteract measures \cite{TanLuHill15,WangYuLiLu16}.

Scrambling is one of most efficient and simple encryption methods \cite{GOLDBURG:scramblers:JSAC1993,KakJayant77}.
Although reference \cite{Lcq:Optimal:SP11} presents optimal quantitative cryptanalysis of any position permutation-only encryption scheme
against known/chosen-plaintext attack, they are still widely used in the design of multimedia encryption schemes. As size of speech and video data is
much larger than that of image data, only some important parameters are selected to be encrypted \cite{WangHempel10,Zeng:VideoScrambling:IEEETMM2003}.
To evaluate their importance degree, various criterions were proposed, such as sensible influence \cite{Servetti02,Zhu:assessment:CASVT2015}, and decoding importance \cite{WuKuo05}. In the field of multimedia secure communication, most efforts were spent on software implementation, only few works touched
hardware implementation of multimedia encryption scheme \cite{LinYuLuCaiChen15}.

In this paper, we propose a novel TCP-based multicast scheme for multiuser speech wireless communication, and
implement it in ARM hardware platform. The scheme is designed for scenario of wireless communication where
the balancing point between sufficient security level and low computation load is expected. The kernel of the scheme is to
compress original speech data input from Audio Codec based on the adaptive differential pulse code modulation (ADPCM) algorithm designed by Interactive Multimedia Association (IMA) \cite{Cummiskey73}. Then, it encrypts the compressed data with the following three levels of chaotic encryption: scrambling 1-bit element in each 1-byte data with a 6-D chaotic map; scrambling position of every 1-byte data with a 7-D chaotic map; encrypting compressed speech data
with a chaotic stream cipher multiple times. The encryption kernel is cascaded with a multicast method connecting sender point to receiver point
with a way of point-to-multiple. Security performance of the proposed system is verified from various aspects: NIST test, statistical analysis, sensitivity
with respect to change of secrete key and size estimation of secret key space. Compared with video encryption scheme proposed in \cite{LinYuLuCaiChen15}, this scheme still run sufficient fast as size of speech data is much smaller than that of video data of the same playing time.

The remainder of this paper is organized as follows. In Sec.~\ref{sec:mcmu}, a brief introduction is given to multicast-multiuser wireless communication system. Section~\ref{sec:DataAcquire} presents basic principle of speech compression. Section~\ref{sec:scrambling} proposes a discrete-time chaotic system without any degeneration and uses it to implement secure and efficient speech communication scheme. Section~\ref{sec:streamcipher} designs a multi-round chaotic stream cipher for speech communication. Section~\ref{sec:multicast} further discusses cascading the proposed speech communication in the multicast environment for multi-user wireless communication. Section~\ref{sec:performance} analyzes security performance of the whole secure communication system. Finally, the last section concludes the paper.

\section{Multicast-multiuser wireless communication system}
\label{sec:mcmu}

\begin{figure}[!htbp]
\centering
\includegraphics[width=\figwidth]{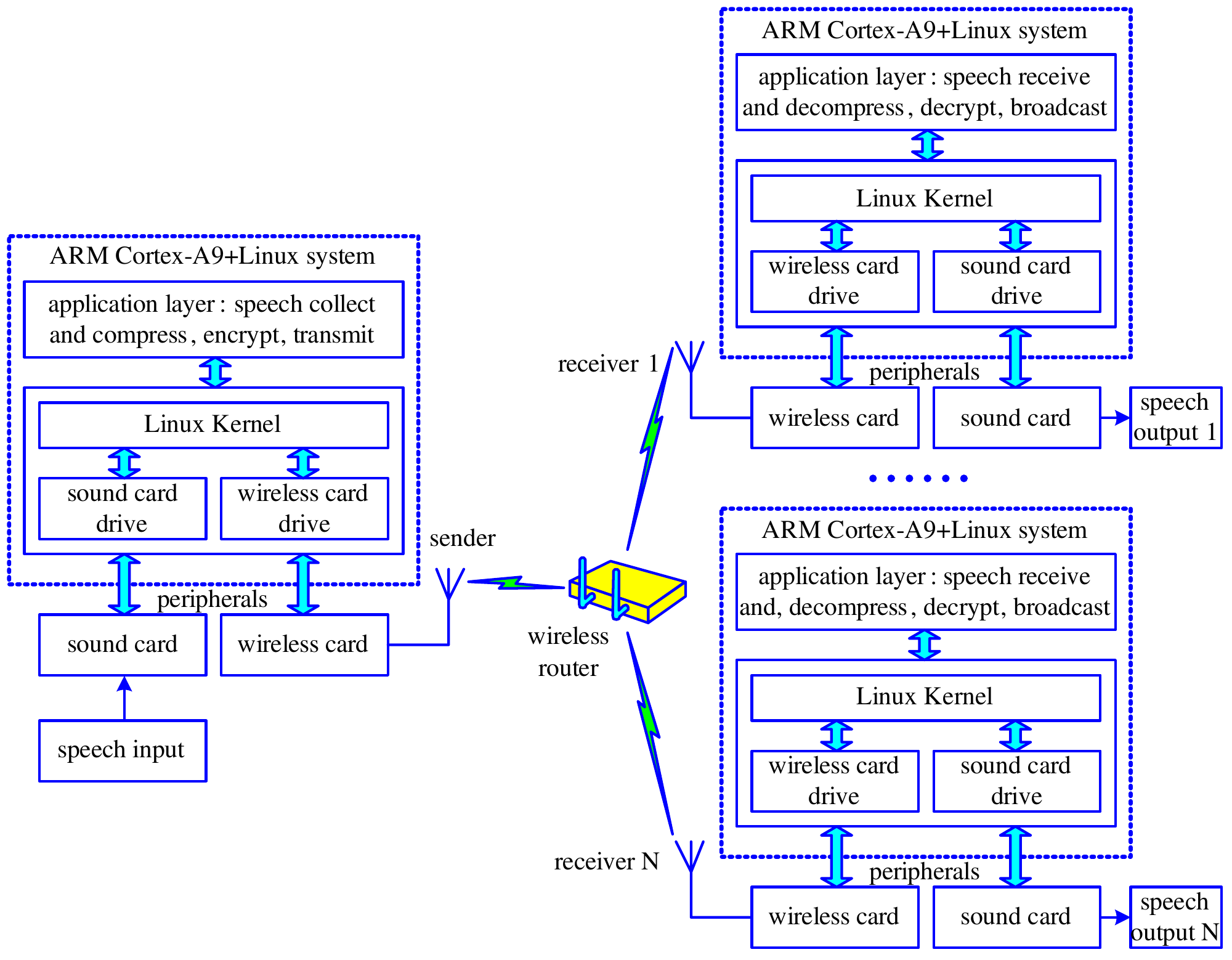}
\caption{The function diagram of the ARM-based multicast-multiuser communication system.}
\label{fig:functionMCMU}
\end{figure}

The function diagram of the ARM-based multicast-multiuser communication system is shown in
Fig.~\ref{fig:functionMCMU}, where both the sender and the receiver use the same development board
ARM9Tiny4412. Its hardware part includes the following key components: four-core Cortex-A9 processing unit
of dominant frequency 1.5GHz, sound card, wireless network card, 1GB DDR3 SDRAM (32-bit data bus), 4GB Flash, and USB port.
The software system can be partitioned into the driver layer and application layer. The former mainly includes sound card driver, network card driver, the Linux kernel and so on.
The latter includes adaptive speech acquisition, speech compression, three-level chaotic encryption, the sender of encrypted speech signal, the receiver for encrypted speech signal, three-level chaotic decryption, speech adaptive decompression, and speech broadcaster.

\begin{figure}[!htbp]
\centering
\includegraphics[width=0.9\figwidth]{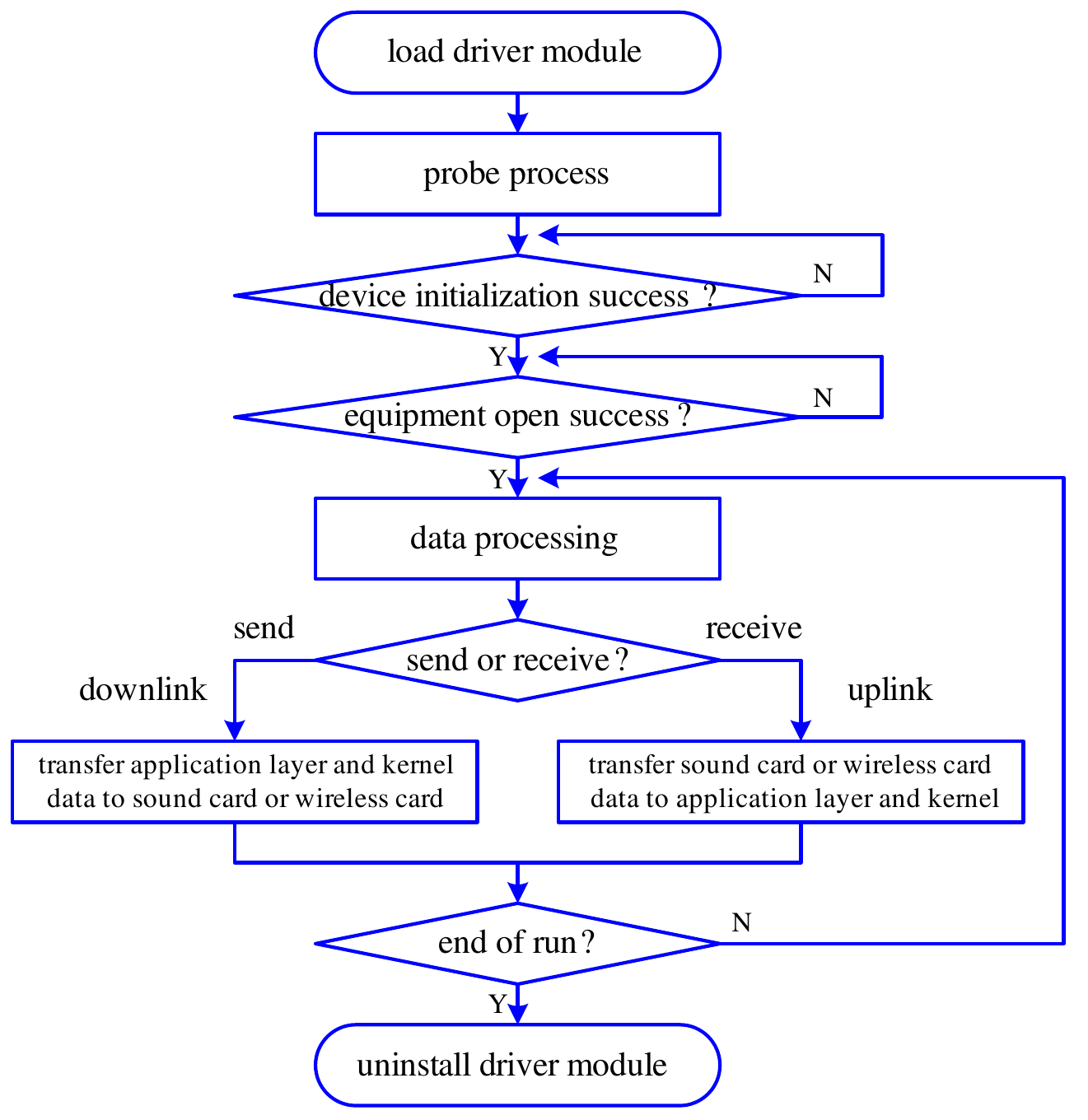}
\caption{Flow chart of uplink/downlink communication.}
\label{fig:chartcommunication}
\end{figure}

As shown in Fig.~\ref{fig:chartcommunication}, the ARM-based multicast-multiuser communication system owns the following features. In the multicast architecture based on ARM+linux embedded devices, a multicast-multiuser chaotic map-based speech communication is implemented. In the uplink communication of both the sending end and receiving end, the received data is sent to the application layer through the sound card, network card and Linux kernel. In the downlink communication, the data is sent to the sound card and network card by them also. All the receivers can receive multicast data via WIFI signal if they registered the multicast membership. When the electricity of ARM is turned on, its own WIFI connection program is run first \cite{Ferro05}. Once any WIFI signal is detected, WIFI is connected successfully by entering the right account number and password.

\section{Design of data acquisition and compression scheme}
\label{sec:DataAcquire}

Redundancy existing in raw message is very useful for error detection and correction in communication. However, it is very harmful in cryptology. As pointed out by Shannon, redundancy provides basis for cryptanalysis, which can be
represented as a formula $U=H/D$, where $U$, $H$ and $D$ denote unicity distance, entropy of secret key, and redundancy degree, respectively.
It means that there is sole solution for deciphering an encryption scheme when the number of ciphertext characters intercepted by the cryptanalysts is greater than the unicity distance. Otherwise, there is multiple solutions and the complexity of the deciphering is very large. From such viewpoint, one can see
that compressing speech data before encryption is helpful for improving security performance of the whole secure communication system.

Every 16-bit digital speech data input from the sound card is compressed into 4-bit data through IMA-ADPCM, whose most significant bit is the sign bit. Since the C language does not support the data types of 4 bits, two variables of char type are used to represent one compressed variable. Concretely, the PCM data
is compressed and stored into buffer in chronological order, and each BYTE space store two compressed data, where the four least significant bits
store the first one and the four most significant bits store the second one.

\begin{figure*}[!htbp]
\center
\includegraphics[width=1.2\figwidth]{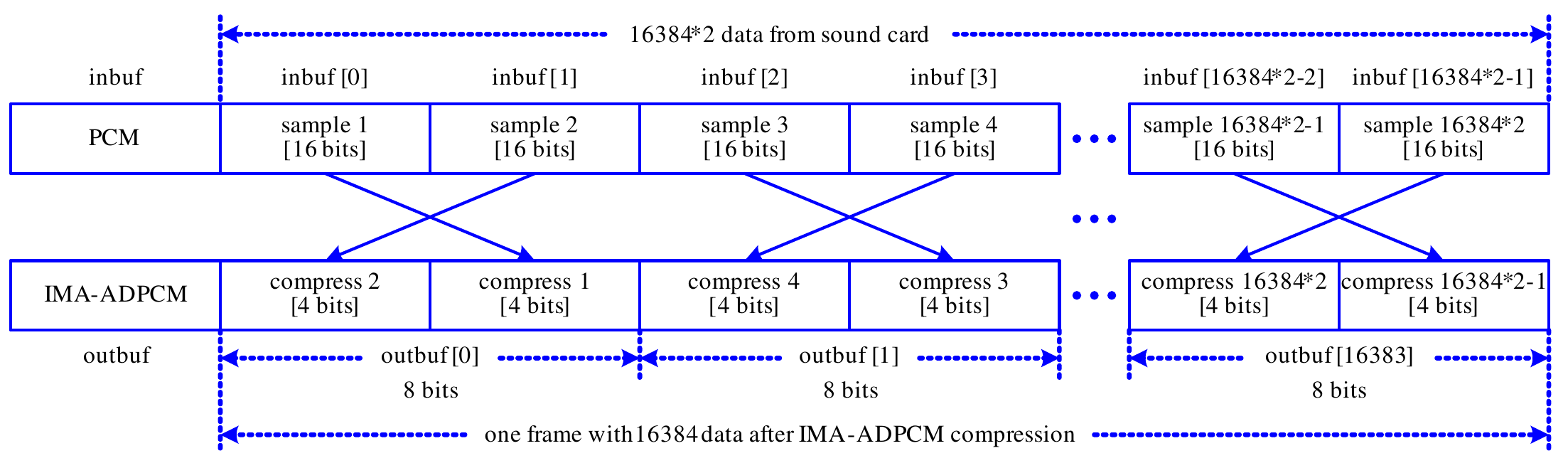}
\caption{Schematic diagram of the speech compression based on IMA-ADPCM.}
\label{fig:SchematicIMA}
\end{figure*}

The schematic of speech compression algorithm based on IMA-ADPCM is shown in Fig.~\ref{fig:SchematicIMA}. First, the 32768 16-bit data collected by sound card are stored into inbuf, and then compressed into 32768 4-bit data by IMA-ADPCM algorithm. Every two adjacent 4-bit data are combined into one 8-bit data.
Finally, 16384 8-bit compressed data are stored into outbuf as one frame.

\section{Design of scrambling scheme for compressed speech data}
\label{sec:scrambling}

This section discusses the design of scrambling scheme for compressed speech data. First, position of every bit of the compressed speech data is permuted randomly with a pseudo-random number sequence (PRNS) generated by a 6-D chaotic map. Then, the scrambled data is further permuted with a PRNS generated by a 7-D chaotic map in the level of byte.

\subsection{Construction of an $n$-dimensional discrete-time chaotic system}

Like \cite{LinYuLuCaiChen15}, we adopt the idea of Chen-Lai chaotification scheme in \cite{YuLuChen13}
to construct an $n$-dimensional discrete-time chaotic system, which can be generally presented as
\begin{equation}
\left\{ \!\!\!\begin{array}{l}
{x_{1,k + 1}} \!=\! \bmod ({A_{11}}{x_{1,k}} \!+\! {A_{12}}{x_{2,k}} \!+\!  \cdots  + {A_{1n}}{x_{n,k}},{\rm{ }}{{\rm{2}}^{{N_1}}}),\\
{x_{2,k + 1}} \!= \!\bmod ({A_{21}}{x_{1,k}} \!+\! {A_{22}}{x_{2,k}} \!+\!  \cdots \! +\! {A_{2n}}{x_{n,k}},{\rm{ }}{{\rm{2}}^{{N_2}}}),\\
 \cdots \\
{x_{n,k + 1}} \!=\! \bmod ({A_{n1}}{x_{1,k}} \!+\! {A_{n2}}{x_{2,k}} \!+\!  \cdots  \!+\! {A_{nn}}{x_{n,k}},{\rm{ }}{{\rm{2}}^{{N_n}}}),
\end{array} \right.
\label{eq:GeneralForm}
\end{equation}
where $\bmod(m, n)=m-n\cdot\lfloor \frac{m}{n} \rfloor$, and $\lfloor x\rfloor$ gives the largest integer less than or equal to $x$.
In the proposed secure communication system, we set $N_1=N_2=\cdots =N_n=N_0$, and $N=2^{N_1}\cdot 2^{N_2}\dots \cdot 2^{N_n}=2^{nN_0}$.

The coefficients $A_{ij}$ in system~(\ref{eq:GeneralForm}) are derived from matrix
\begin{equation}
\mathbf{A}=
\begin{bmatrix}
A_{11} & A_{12} & \ldots & A_{1n} \\
A_{21} & A_{22} & \ldots & A_{2n} \\
\vdots & \vdots & \ddots & \vdots \\
A_{n1} & A_{n2} & \ldots & A_{nn}
\end{bmatrix}
=\prod\limits_{i = 1}^{n - 1} {\left( {\prod\limits_{j = i + 1}^n {\mathbf{T}_{ij}} } \right),}
\label{eq:matrixmultiply}
\end{equation}
where
$\mathbf{T}_{ij}$ is given by
\begin{equation}
\setlength{\arraycolsep}{2pt}
\mathbf{T}_{ij} = \begin{pmatrix}
1        &0           &\cdots   &               &  &   &  \\
0        &1           &\cdots   &\mathbf{0}     &  &0  &\\
\vdots   &\vdots      &\ddots   &               &  &   &  \\
         &\mathbf{0}  &         &\begin{matrix}\begin{smallmatrix}
{A_{i,\;i}^{(11)}}&0& \cdots   &0& \cdots &0&{A_{i,\;j}^{(12)}}\\
0&1& \cdots &0& \cdots &0&0\\
 \vdots & \vdots & \ddots & \vdots & \vdots & \vdots & \vdots \\
0&0& \cdots &1& \cdots &0&0\\
 \vdots & \vdots & \cdots & \vdots & \ddots & \vdots & \vdots \\
0&0& \cdots &{\rm{0}}& \cdots &1&{\rm{0}}\\
{A_{j,\;i}^{(21)}}&{\rm{0}}& \cdots &{\rm{0}}& \cdots &{\rm{0}}&{A_{j,\;j}^{(22)}} \end{smallmatrix}\end{matrix}         &   & \mathbf{0}  & \\
         &          &    &  &\ddots  &\cdots  &\cdots  \\
         &0         &    &\mathbf{0} &\vdots  &1  &0   \\
         &          &    &           &\vdots  &0  &1
\end{pmatrix}.
\label{eq:matrix}
\end{equation}

\begin{algorithm}
\caption{The \texttt{Transform} function}
\begin{algorithmic}[1]
\Function{\texttt{Transform}}{A}
\State $S=0$
\For {$x_{1, k} \leftarrow 0, 2^{N_0}-1$}
\For {$x_{2, k} \leftarrow 0, 2^{N_0}-1$}
$\cdots $
\For {$x_{n, k} \leftarrow 0, 2^{N_0}-1$}
  \State $E(S, 1)=\mod(A_{11}x_{1,k}+A_{12}x_{2,k}+\cdots +A_{1n}x_{n,k}, 2^{N_0})$
   \State $E(S, 2)=\mod(A_{21}x_{1,k}+A_{22}x_{2,k}+\cdots +A_{2n}x_{n,k}, 2^{N_0})$
  \State $\cdots$
    \State $E(S, n)=\mod(A_{n1}x_{1,k}+A_{n2}x_{2,k}+\cdots +A_{nn}x_{n,k}, 2^{N_0})$
    \State $S=S+1$
   \EndFor $\cdots $
    \EndFor
     \EndFor
\State \textbf{return} $E$
\EndFunction
\end{algorithmic}
\label{algo:array}
\end{algorithm}

The mathematical expression of the serial number is
\begin{equation}
S=\sum_{i=1}^{nN_0/2} x_{i, k}\cdot 2^{n\cdot N_0-2\cdot i}.
\label{eq:numberS}
\end{equation}
Utilizing Algorithm~\ref{algo:array}, the corresponding scrambling index can be obtained via
\begin{equation}
E(S)=\sum_{i=1}^{nN_0/2} E(S, \alpha_i)\cdot 2^{n\cdot N_0-2\cdot i},
\label{eq:numberPerS}
\end{equation}
where $\alpha_i\in \{1, 2, \cdots, n\}$, and $\alpha_1\neq \alpha_2\neq \cdots \neq \alpha_n$.

\subsection{Scrambling the compressed speech data in the level of bit}

\begin{figure}[!htbp]
\centering
\includegraphics[width=\figwidth]{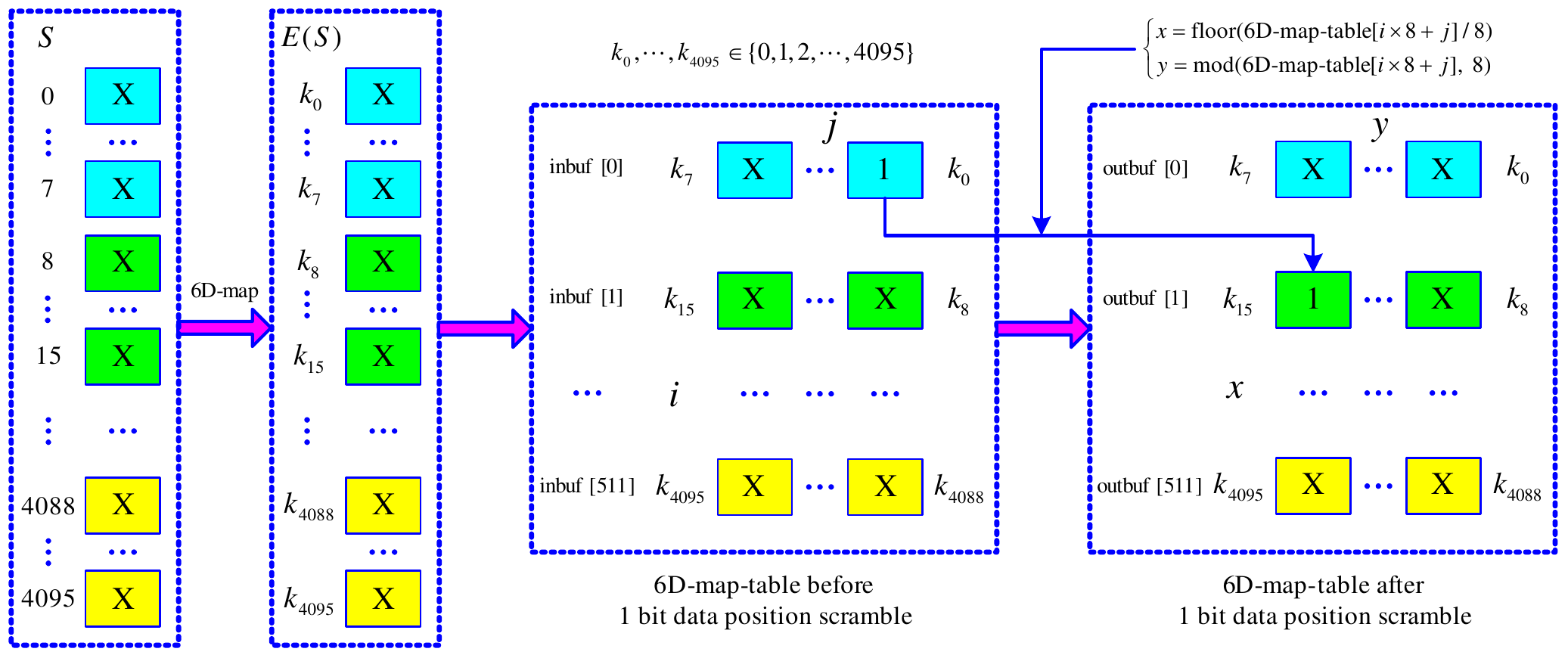}
\caption{Schematic diagram of bit permutation.}
\label{fig:DiagramPermutation}
\end{figure}

\begin{figure}[!htbp]
\centering
\includegraphics[width=0.7\figwidth]{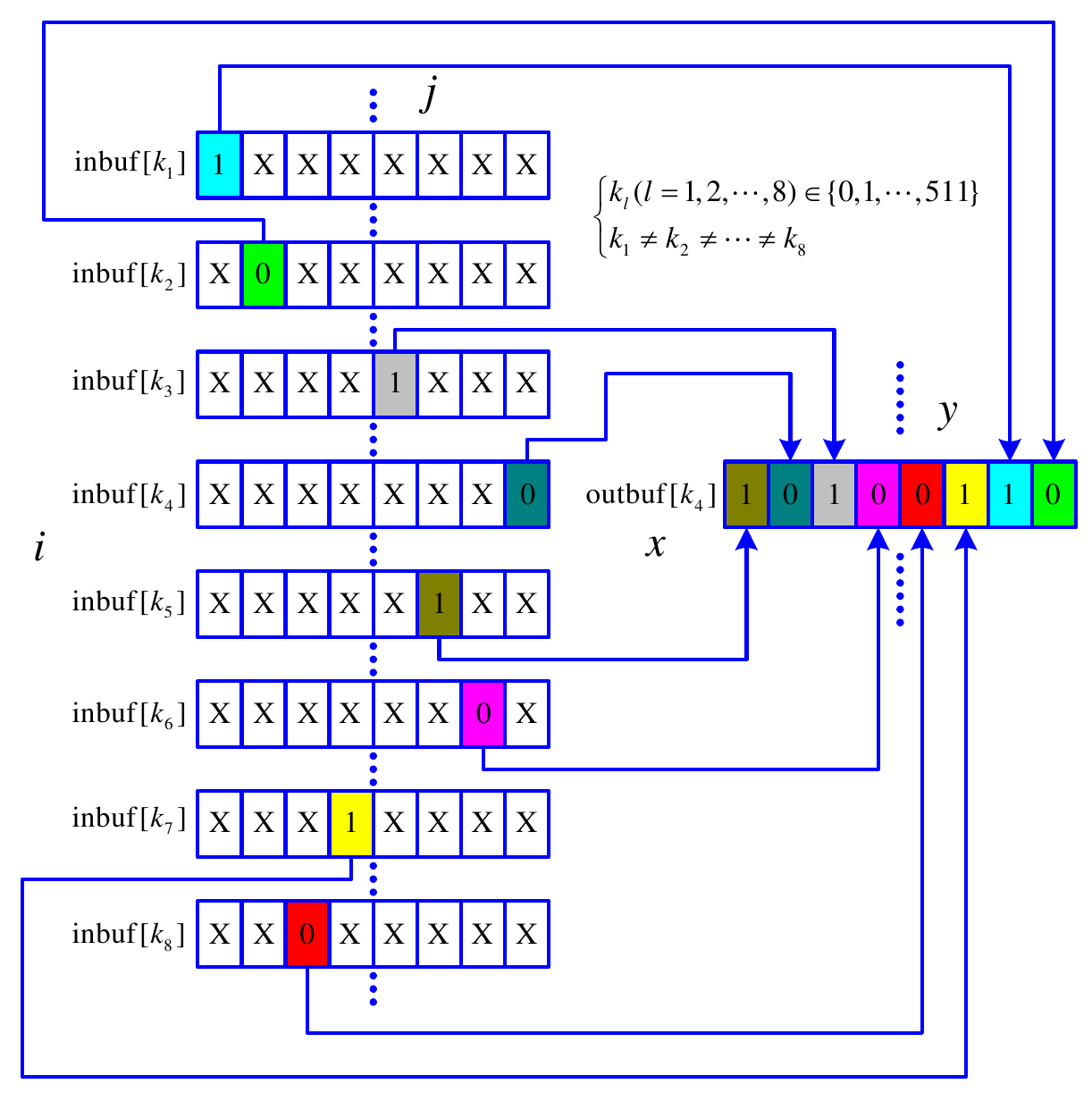}
\caption{Schematic diagram of eight rounds of bit permutation.}
\label{fig:Diagram8rPermutation}
\end{figure}

Set the variables in system~(\ref{eq:GeneralForm}) as $n=6$, $N_0=2$, the number of possible scrambling is
$N=2^{n\cdot N_0}=2^{6\cdot 2}=4096$. Then, position of every bit of the compressed data is permuted randomly with a PRNS generated by the 6-D chaotic map constructed in the above sub-section. As illustrated in Fig.~\ref{fig:SchematicIMA}, each frame of compressed speech data has 16384 bits.
So, it needs to be divided into 32 groups, and each group has 512 bytes. There are $512\cdot 8=4096$ bits of data, so the positions of the 4096 bits of data are scrambled to complete the encryption. After repeating the method 32 times to complete encryption of one frame of compressed speech data, the diagram of scrambling 1-bit data position is shown in Fig.~\ref{fig:DiagramPermutation}. Similarly, the 8-bit data in $\text{inbuf}[k_4]$
becomes $\text{outbuf}[k_4]$ by scrambling 1-bit data position 8 times, which is shown in Fig.~\ref{fig:Diagram8rPermutation}.
The original speech compressed data is temporarily stored in the inbuf, the encrypted data is stored in the outbuf, and all bits of data in
the outbuf are set as zero initially.

According to Fig.~\ref{fig:DiagramPermutation}, the corresponding chaotic encryption algorithm is described as follows:
\begin{itemize}
\item \textit{Step 1}: According to Eq.~(\ref{eq:numberS}), Eq.~(\ref{eq:numberPerS}) and Fig.~\ref{fig:DiagramPermutation}, the value of $S$ is mapped to
$E(S)$, and the corresponding coordinate is $(i, j)$;

\item \textit{Step 2}: Extract 1-bit data unit corresponding to the $i$-th row and the $j$-th column in inbuf as
\[
(\text{inbuf}[i]>>j)\& 1 \in\{0, 1\}.
\]

\item \textit{Step 3}: The positions of data are then scrambled by 6-D chaotic mapping table, and the coordinates are mapped to
\begin{equation}
\left\{
\begin{IEEEeqnarraybox}[
\IEEEeqnarraystrutmode
\IEEEeqnarraystrutsizeadd{2pt}
{2pt}
][c]{rCl}
x & = & \lfloor \mbox{6-D-map-table}[i\cdot 8+j]/8 \rfloor,\\
y & = & \bmod( \mbox{6-D-map-table}[i\cdot 8+j], 8),
\end{IEEEeqnarraybox}
\right.
\label{eq:mapp}
\end{equation}

\item \textit{Step 4}: Replace 1-bit data in the corresponding unit from $(i, j)$ to $(x, y)$ via
\[
\text{Outbuf}[x]=(( (\text{Inbuf}[i]\gg j)\& 1)\ll y)\vee \text{Outbuf}[x],
\]
where the symbol ``$\gg j$" denotes right shift of $j$ bits, ``$\ll y$" denotes left shift of $y$ bits, ``$\&$" denotes the bitand operation, ``$\vee$" denotes the bitwise OR operation, and ``$\bmod$" represents modulo operation.
\end{itemize}

Set the parameters in Eq.~(\ref{eq:matrix}), $A_{i, i}^{(11)}=A^{(11)}=1$, $A_{i, j}^{(12)}=A^{(12)}=1$, $A_{j, i}^{(21)}=A^{(21)}=1$,
$A_{j, j}^{(22)}=A^{(22)}=2$, $n=6$, then one has
\begin{equation}
\mathbf{A}_6=\prod\limits_{i = 1}^{5} {\left( {\prod\limits_{j = i+1}^6 {\mathbf{T}_{ij}} } \right)=\left(
                                                                                                      \begin{array}{cccccc}
                                                                                                        1 & 5 & 18 & 52 & 121 & 197\\
                                                                                                        1 & 6 & 22 & 64 & 149 & 242 \\
                                                                                                        1 & 5 & 19 & 55 & 128 & 208 \\
                                                                                                        1 & 4 & 13 & 38 & 88 & 144 \\
                                                                                                        1 & 3 & 8  & 20 & 48 & 80 \\
                                                                                                        1 & 2 & 4  & 8  & 16 & 32 \\
                                                                                                      \end{array}
                                                                                                    \right).
}
\label{eq:examplemat}
\end{equation}

A concrete example is provided to illustrate the algorithm for scrambling of 1-bit data position:
\begin{itemize}
\item According to Eq.~(\ref{eq:numberS}), Eq.~(\ref{eq:numberPerS}) and Fig.~\ref{fig:DiagramPermutation},
when $S=14$, $E(S)=1792$, then the corresponding coordinate $(i, j)=(1, 6)$.

\item Extract 1-bit data unit corresponding to the $i=1$-th row and the $j=6$-th column in inbuf as
\[
(\text{inbuf}[1]>>6)\& 1=0.
\]

\item The positions of data are then scrambled by 6-D chaotic mapping table, and the coordinates are mapped to
\begin{equation*}
\left\{
\begin{IEEEeqnarraybox}[
\IEEEeqnarraystrutmode
\IEEEeqnarraystrutsizeadd{2pt}
{2pt}
][c]{rCl}
x & = & \lfloor \mbox{6-D-map-table}[1\cdot 8+6]/8 \rfloor=\lfloor 1792/8 \rfloor=224,\\
y & = & \bmod( \mbox{6-D-map-table}[1\cdot 8+6], 8)\\
  & = & \bmod(1792, 8)=0.
\end{IEEEeqnarraybox}
\right.
\end{equation*}

\item The 1-bit data in the corresponding unit is moved from $(i, j)=(1, 6)$ to $(x, y)=(224, 0)$ via
\[
\mbox{Outbuf}[224]=(( (\mbox{Inbuf}[1]\gg 6)\& 1)\ll 0)\vee \mbox{Outbuf}[224].
\]
\end{itemize}

\begin{figure*}[!htbp]
\center
\includegraphics[width=1.2\figwidth]{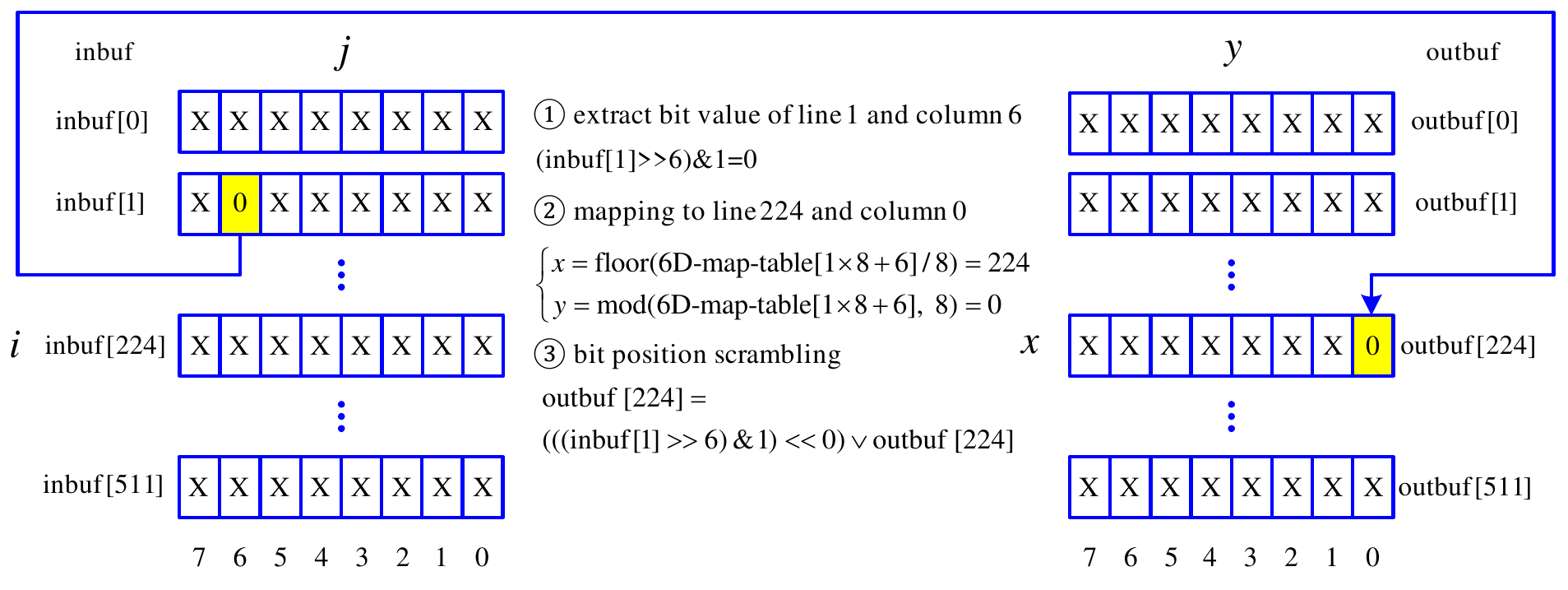}
\caption{The schematic diagram of scrambling one typical element.}
\label{fig:DigramOnePixel}
\end{figure*}

Finally, all the 0's in $(i, j)=(1, 6)$ are replaced by the corresponding unit in $(x, y)=(224, 0)$. The rest cases follow the rule shown in Fig.~\ref{fig:DigramOnePixel} similarly.

To decrypt the received messages, two cases must be distinguished: communication with matched parameters and that with mismatched parameters. Taking the order of 15 matrices in Eq.~(\ref{eq:examplemat}) as the key, the analysis results with matched parameters and that with mismatched parameters are shown as follows:

1) With matched parameters, the permutation matrix at the receiving end is still the same as that shown in Eq.~(\ref{eq:examplemat}).
When $(x, y)=(224, 0)$, one has $(8x+y)=(8\cdot 224+0)=1792$. After the decryption, 0's can be replaced back to the correct position
$(i, j)=(1, 6)$. Similarly, the other encrypted information can be correctly decrypted at the receiving end.

2) Assume that the original arrangement is
\begin{equation*}
\mathbf{T}_{12}\mathbf{T}_{13}\mathbf{T}_{14}\mathbf{T}_{15}\mathbf{T}_{16}\mathbf{T}_{23}\mathbf{T}_{24}\mathbf{T}_{25}\mathbf{T}_{26}
\mathbf{T}_{34}\mathbf{T}_{35}\mathbf{T}_{36}\mathbf{T}_{45}\mathbf{T}_{46}\mathbf{T}_{56},
\end{equation*}
$\mathbf{T}_{12}$ and $\mathbf{T}_{34}$ are swapped by the mismatched parameters, the transformation matrix become
\begin{equation*}
\mathbf{A}_6=\left(
\begin{array}{cccccc}
  5  & 9  & 11 & 29 & 68 & 113\\
  1  & 2  & 3 & 8 & 19 &  31 \\
  11 & 20 & 24 & 65 & 152 & 252 \\
  16 & 29 & 33 & 91 & 212 & 352 \\
  4  & 7 & 8  & 20 & 48 & 80 \\
  3  & 5 & 4  & 8  & 16 & 32 \\
\end{array}
\right).
\end{equation*}

As $8x+y=8\cdot 224+0=1792$, 0's is replaced by the decryption process to a wrong position, $(i, j)=(0, 1)$ , instead of right position $(i, j)=(1, 6)$.
So, the encrypted information cannot be correctly decrypted at the receiving end.

\subsection{Scrambling the compressed speech data in the level of BYTE}

\begin{figure}[!htbp]
\centering
\includegraphics[width=\figwidth]{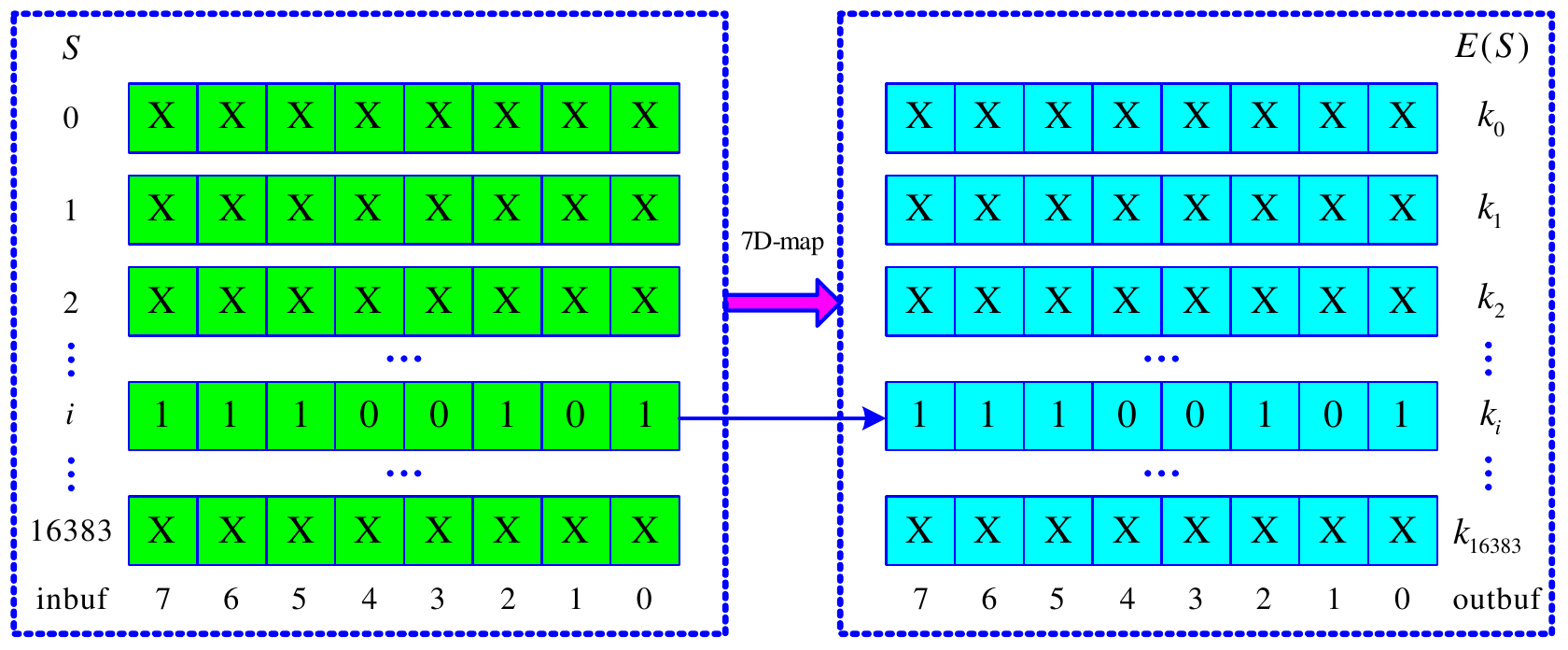}
\caption{Schematic diagram of scrambling and anti-scrambling of 8 bits speech compressed data.}
\label{fig:permutation}
\end{figure}

In the second-level of encryption, let $n=7$, $N_0=2$, the number of the position scrambling is $N=2^{n\cdot N_0}=2^{7\cdot 2}=16384$.
The data is further permuted in the level of byte with a PRNS generated by a 7-D chaotic map. Positions of frames of size $1\times 16384$ are then scrambled as shown in Fig.~\ref{fig:permutation}, where the original speech compressed data is temporarily stored in the inbuf, the encrypted data is temporarily stored in the outbuf,
where $k_i\in \{0, 1, 2, \cdots, 16383\}$ and $k_0\neq k_1\neq k_2\neq \cdots \neq k_{16383}$.

When $n=7$, the sequence number $S$ of the speech compression data before encryption is
\begin{equation}
S=\sum_{i=1}^7 x_{i, k}\cdot 2^{14-2i}.
\label{eq:indexS7}
\end{equation}
The corresponding sequence number $S$ of the speech compression data after encryption is
\begin{equation}
E(S)=\sum_{i=1}^7E(S, \alpha_i)\cdot 2^{14-2i},
\label{eq:indexiS7}
\end{equation}
where
$\alpha_i\in \{1, 2, \cdots, 7\}$, $\alpha_1\neq \alpha_2 \cdots \neq \alpha_7$.

Setting coefficients in Eq.~(\ref{eq:matrix}),
$A_{i, i}^{(11)}=A^{(11)}=1$, $A_{i, j}^{(12)}=A^{(12)}=1$, $A_{j, i}^{(21)}=A^{(21)}=1$,
$A_{j, j}^{(22)}=A^{(22)}=2$, $n=7$, one has
\begin{IEEEeqnarray}{rCl}
\mathbf{A}_7 & = & \prod\limits_{i = 1}^{6} \left( {\prod\limits_{j = i+1}^7 {\mathbf{T}_{ij}} } \right)  \label{eq:examplematn7}\\
& = & \left(
\begin{array}{ccccccc}
    1 & 6 & 25 & 84 & 237 & 550 & 903 \\
    1 & 7 & 30 & 102& 289 & 671 & 1100 \\
    1 & 6 & 26 & 88 & 249 & 578 & 948 \\
    1 & 5 & 19 & 63 & 176 & 408 & 672 \\
    1 & 4 & 13 & 38 & 104 & 240 & 400 \\
    1 & 3 & 8  & 20 & 48 & 112 & 192 \\
    1 & 2 & 4  & 8  & 16 & 32 & 64 \\
  \end{array}
  \right).\nonumber
\end{IEEEeqnarray}

A concrete example is illustrated the algorithm for scrambling position of every byte.
Let $S=i=100$, $E(S)=E(i)=k_i=3296$. The position $S=100$ for 8-bit data ``11100101" is replaced by $E(S)=3296$.

Taking the order of 21 matrices in Eq.~(\ref{eq:examplematn7}) as the key, the analysis results with matched parameters and that with mismatched parameters are shown as follows.

1) With matched parameters, the permutation matrix at the receiving end is still that shown in Eq.~(\ref{eq:examplematn7}).
Given $E(S)=3296$, one can get $S=100$. After the decryption process, the 8-bit data is replaced back to the right position.
The other cases run in the similar way. So, the encrypted speech data can be decrypted successfully.

2) Assume the original arrangement is
\begin{multline*}
\mathbf{T}_{12}\mathbf{T}_{13}\mathbf{T}_{14}\mathbf{T}_{15}\mathbf{T}_{16}\mathbf{T}_{17}\mathbf{T}_{23}\mathbf{T}_{24}\mathbf{T}_{25}\mathbf{T}_{26}
\mathbf{T}_{27}\mathbf{T}_{34}\mathbf{T}_{35}\mathbf{T}_{36}\mathbf{T}_{37}\mathbf{T}_{45}\mathbf{T}_{46}\mathbf{T}_{47}\mathbf{T}_{56}\mathbf{T}_{57}
\mathbf{T}_{67},
\end{multline*}
$\mathbf{T}_{12}$ and $\mathbf{T}_{45}$ are swapped by mismatched parameters, the transformation matrix becomes
\[
A_7=\left(
  \begin{array}{ccccccc}
    6 & 11 & 20 & 53 & 138 & 322 & 539 \\
    1 & 2 & 5 & 14 & 37 & 87 & 144 \\
    7  & 13 & 26 & 70 & 183 & 428 & 714 \\
    11 & 20 & 32 & 81 & 212 & 492 & 828 \\
    16 & 29 & 45 & 111 & 292 & 676 & 1140 \\
    4  & 7 & 8    & 20 & 48 & 112 & 192 \\
    3 & 5 & 4 & 8 & 16 & 32 & 64 \\
  \end{array}
\right).
\]

Setting the above matrix and Eq.~(\ref{eq:indexS7}), (\ref{eq:indexiS7}), $S=491$, one can get $E(S)=3296$. After the decryption, the byte is replaced to the wrong position of index $S=491$ instead of index $S=100$. So, the encrypted information cannot be correctly decrypted at the receiving end.

\section{Encrypting compressed speech data by multi-round stream cipher}
\label{sec:streamcipher}

This section discusses design of the third level of chaotic encryption scheme for the compressed speech data.
First, we propose a general design method for discrete chaotic systems without degeneration of positive Lyapunov exponent.
Then, we use 3-D version of the chaotic systems to encrypt 8-bit compressed speech data multiple times.

\subsection{Design of discrete time chaotic system without degeneration of positive Lyapunov exponents}

The relation between random statistical characteristic of a chaotic system and Kolmogorov-Sinai entropy can be represented as
\[h_{KS}=\sum_iLE_i^+,\]
where $LE_i^+$ is the number of positive Lyapunov exponents \cite{YuLuChen13}.
As the larger value of $h_{KS}$ means the better statistical properties of randomness, a chaotic system owns better random statistical performance
if it has more and larger positive Lyapunov exponents. In our design of discrete time chaotic system, we try to make the number and the quantity of
its positive Lyapunov exponents both reach maximum. The concrete approaches can be described as follows.

{\bf{Step 1}}: The general form of a nominal system with asymptotically stable origin is
\begin{equation}
\mathbf{x}(k+1)=\mathbf{C}\mathbf{x}(k),
\label{eq:NominalSystem}
\end{equation}
where
$
\mathbf{C}=\left(
\begin{array}{ccc}
c_{11} & \cdots & c_{1n} \\
\vdots & \ddots & \vdots \\
c_{n1} & \cdots & c_{nn} \\
\end{array}
\right)$ is a constant matrix, and all characteristic roots of the nominal system are located inside an unit circle.

{\bf{Step 2}}: To control the nominal system (\ref{eq:NominalSystem}) effectively, transform it as
\begin{equation}
\mathbf{x}(k+1)=\mathbf{A}\mathbf{x}(k),
\label{eq:NominalSystem2}
\end{equation}
where
\begin{equation*}
\mathbf{A}=\mathbf{P}\mathbf{C}\mathbf{P}^{-1},
\end{equation*}
and $\mathbf{P}$ is a nonsingular matrix. Due to the properties of similarity transformation, the two systems, (\ref{eq:NominalSystem}) and (\ref{eq:NominalSystem2}), have the same characteristic roots and stability performance.

{\bf{Step 3}}: Control the system (\ref{eq:NominalSystem2}) with the strategy of anticontrol proposed in \cite{YuLuChen13} and get a uniformly bounded system
\begin{equation}
\mathbf{x}(k+1)=\mathbf{A}\mathbf{x}(k)+\mathbf{B}\mathbf{g}(\mathbf{\sigma} \mathbf{x}(k), \mathbf{\varepsilon}),
\label{eq:ControlledSystem}
\end{equation}
where $\mathbf{g}(\mathbf{\sigma} \mathbf{x}(k), \mathbf{\varepsilon})$ is a uniformly bounded controller and $\mathbf{B}$ is a control matrix.

{\bf{Step 4}}: Exert pole assignment on the system (\ref{eq:ControlledSystem}) by adjusting $\mathbf{B}$, $\mathbf{\sigma}$ and $\mathbf{\varepsilon}$ and
make the number of its positive Lyapunov exponents $L$ reach maximum and the quantities can be sufficiently large, i.e. the system (\ref{eq:ControlledSystem})
reach the state without degeneration.

To illustrate performance of the above method, we present a concrete example of 3-D system,
\begin{equation}
\left\{
\begin{IEEEeqnarraybox}[
\IEEEeqnarraystrutmode
\IEEEeqnarraystrutsizeadd{2pt}
{2pt}
][c]{rCl}
x_1(k+1) & = & a_{11}x_1(k)+a_{12}x_2(k)+a_{13}x_3(k) \triangleq f_1(\cdot),\\
x_2(k+1) & = & a_{21}x_1(k)+a_{22}x_2(k)+a_{23}x_3(k)   \triangleq f_2(\cdot),\\
x_3(k+1) & = & a_{31}x_1(k)+a_{32}x_2(k)+a_{33}x_3(k)\\
               &  & {\;\;}    +\varepsilon\sin(\sigma x_1(k))\triangleq f_3(\cdot),
\end{IEEEeqnarraybox}
\right.
\label{eq:example}
\end{equation}
where
\[
\mathbf{A}=\begin{pmatrix}
   a_{11} & a_{12} & a_{13}  \\
   a_{21} & a_{22} & a_{23}  \\
   a_{31} & a_{32} & a_{33}
\end{pmatrix}
=
\begin{pmatrix}
   0.205  & -0.595 & 0.265 \\
   -0.265 & -0.125 & 0.595 \\
   0.33 & -0.33  & 0.47
\end{pmatrix},
\]
and $\sigma=6.6667\varepsilon/10^{-4}$.

\begin{figure}[!htbp]
\centering
\includegraphics[width=0.8\figwidth]{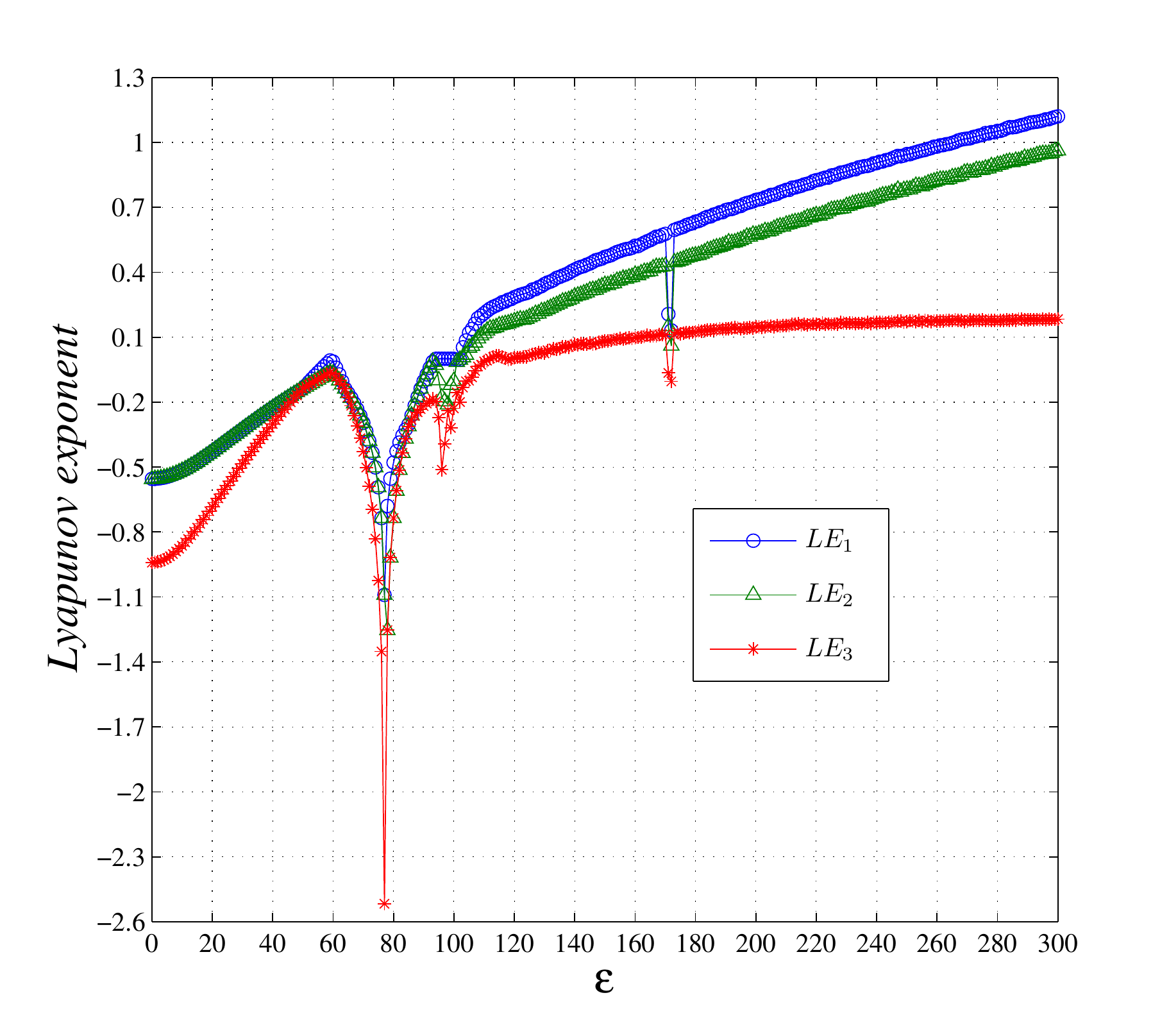}
\caption{Lyapunov spectrum with parameter $\epsilon$ and $\sigma$.}
\label{fig:Lyapunov}
\end{figure}

\begin{figure}[!htbp]
\center
\begin{minipage}{0.6\figwidth}
\center
\includegraphics[width=0.5\figwidth]{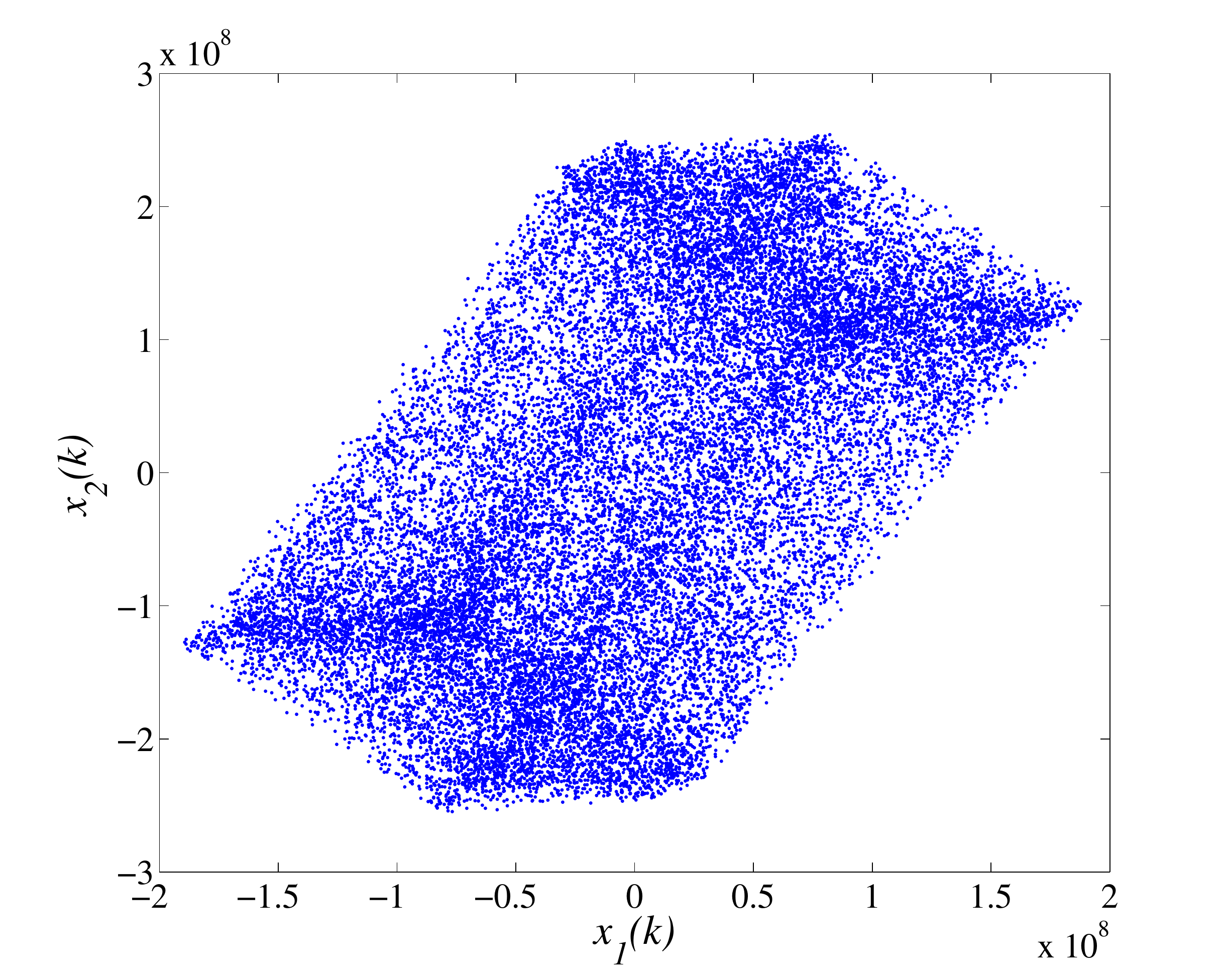}
a)
\end{minipage}
\begin{minipage}{0.6\figwidth}
\center
\includegraphics[width=0.5\figwidth]{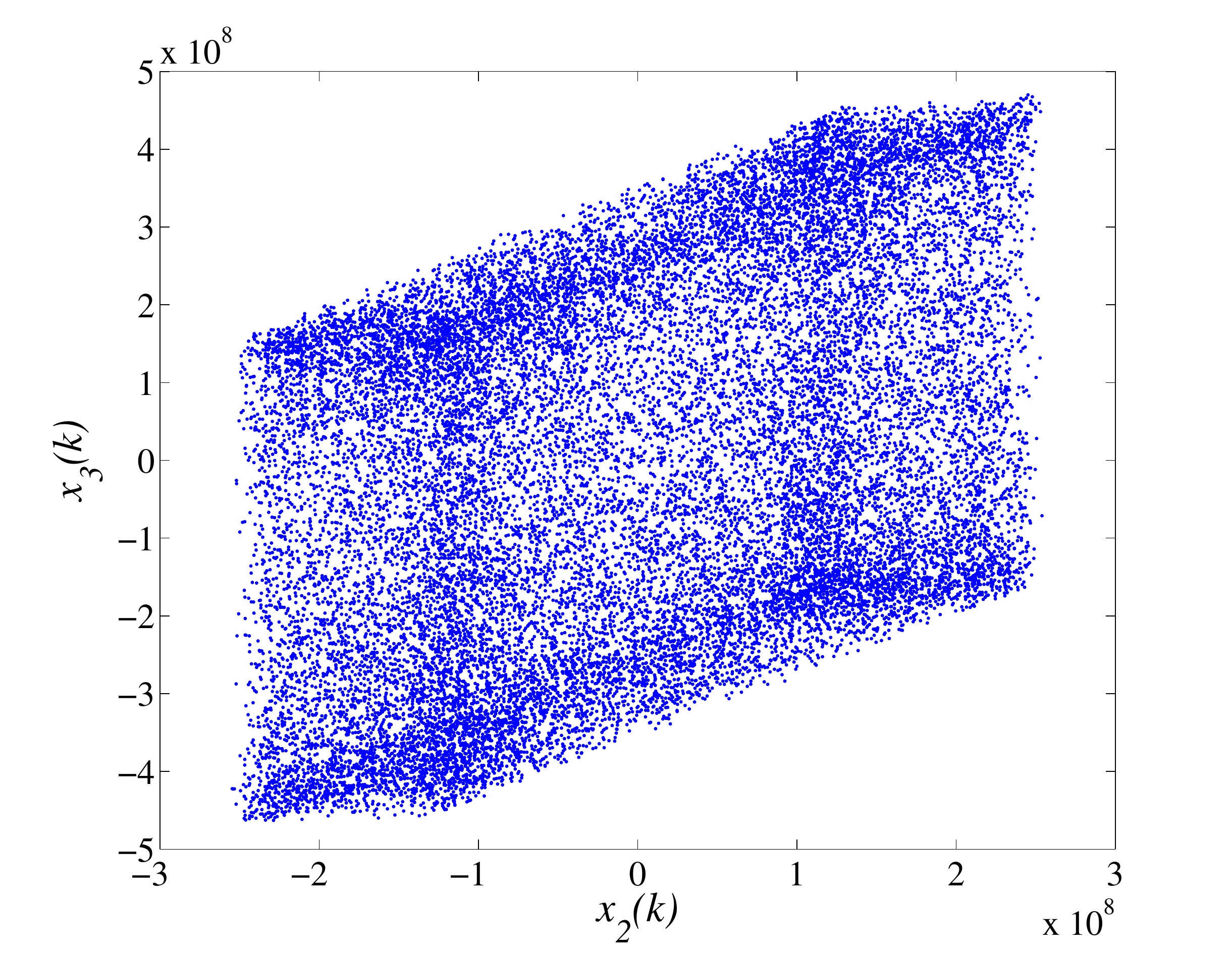}
b)
\end{minipage}
\caption{Phase diagram of a 3-D chaotic attractor: a) $x_1-x_2$ plane; b) $x_2-x_3$ plane.}
\label{fig:phase}
\end{figure}

Figure~\ref{fig:Lyapunov} depicts Lyapunov spectrum of the system~(\ref{eq:example})
for $\varepsilon\in [0, 300]$, where $LE_1$, $LE_2$ and $LE_3$ denote Lyapunov exponents in three dimensions, respectively.
From Fig.~\ref{fig:Lyapunov}, we can see that the Lyapunov exponent becomes
more and more large and be positive in most cases when $\varepsilon>100$. When $\varepsilon=3\cdot 10^8$,
$\sigma=2\cdot 10^5$, one can calculate $LE_1=14.9$, $LE_2=14.8$, $LE_2=0.19$, the corresponding phase diagram is shown
in Fig.~\ref{fig:phase}.

\subsection{The design principle of multi-round stream cipher}

\begin{figure}[!htbp]
\center
\includegraphics[width=0.9\figwidth]{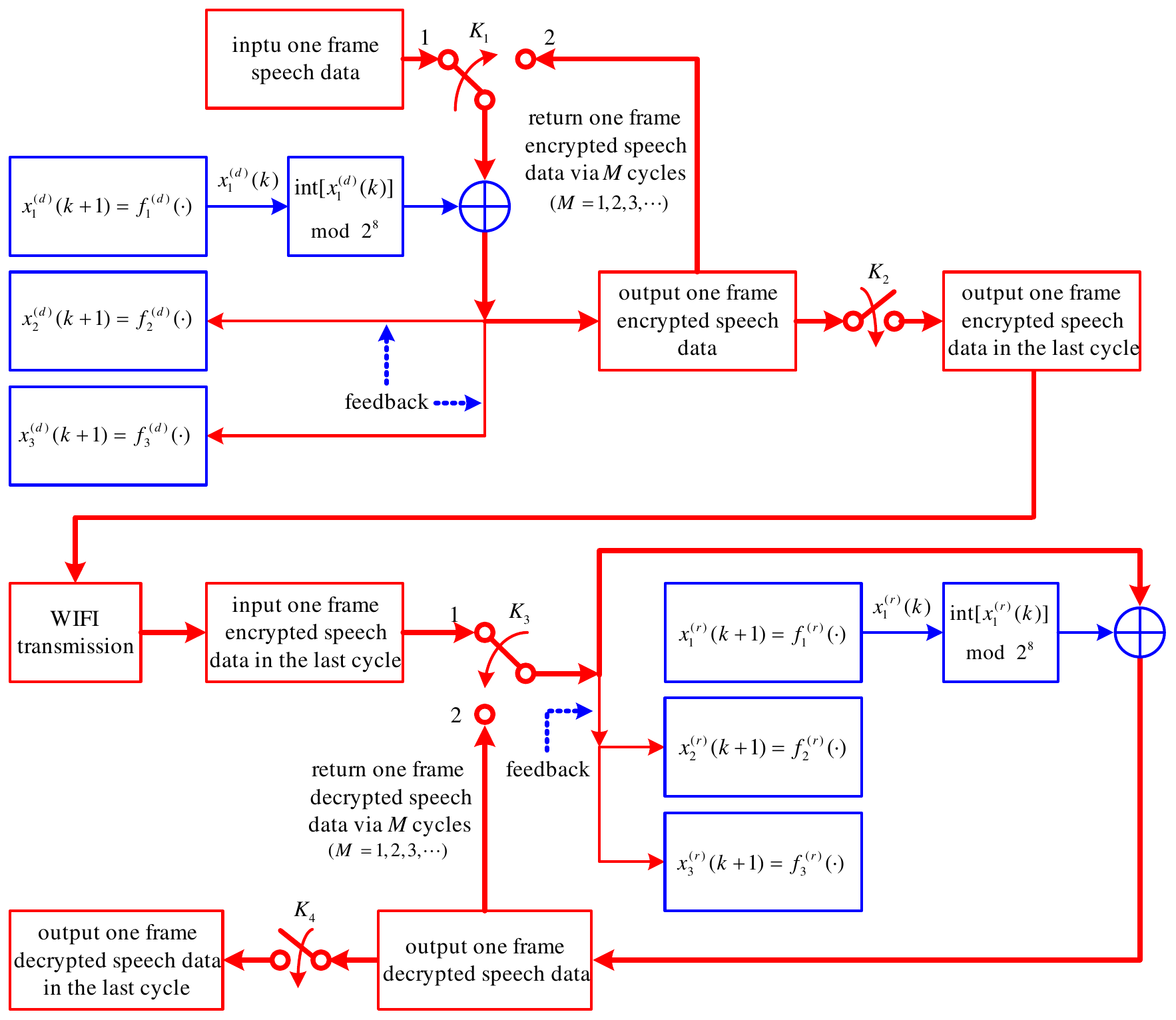}
\caption{Design principle of multi-round stream cipher.}
\label{fig:Principle}
\end{figure}

The design principle of multi-round stream cipher is shown in Fig.~\ref{fig:Principle}.
The iterative equations $x_i^{(d)}(k+1)=f_i^{(d)}(\cdot)$ and $x_i^{(r)}(k+1)=f_i^{(r)}(\cdot)$ ($i=1, 2, 3$) defined in system (\ref{eq:example}), are
implemented with double-precision arithmetic. Functions $\textrm{int}[x_1^{(d)}(k)]$ and $\textrm{int}[x_1^{(r)}(k)]$ transform the variables from double type to 64-bit integer. The modulo function $\bmod(2^8)$ get the eight least significant bits of the 64-bit integer, which will be used for encryption of 8-bit speech signal. To further enhance the ability to resist differential attack, we propose an encryption scheme by $M$-rounds chaotic stream cipher for each frame of the speech data. Security analysis shows that this scheme can greatly enhance the ability to resist differential attack. The whole framework shown in Fig.~\ref{fig:Principle} consists of four parts: encryption scheme based on multi-round of chaotic stream cipher at the sending end, decryption scheme based on multi-round of chaotic stream cipher at the receiving end, channel and the switch function of $K_1$, $K_2$, $K_3$, and $K_4$. The four switch functions can be programmed through C language. The main feature of this scheme is that the $x_1^{(d)}(k)$ at second and third equations of chaotic system in Fig.~\ref{fig:Principle} are replaced by the output feedback through closed-loop feedback (including the speech signal) at the sending end. Then, the final encrypted speech signal after $M$-round encryption is transmitted via WIFI. At the receiving end, $x_1^{(r)}(k)$ at second and third equations of chaotic system are replaced by the received encrypted speech signal after $M$-round encryption. Then, the corresponding $M$-rounds of speech signal decryption makes chaotic systems at the sending end and the receiving end synchronize, i.e., the decryption can be achieved successfully.

The basic operation principle of encryption scheme based on multi-round of chaotic stream cipher shown in Fig.~\ref{fig:Principle} is summarized as follows:

(1) At the sending end, the operation principle of the $M$-round encryption for each frame of the speech data is as follows: First, place $K_1$ at position 1 in Fig.~\ref{fig:Principle} and encrypt a frame of signal. Then, place $K_1$ at position~2 in Fig.~\ref{fig:Principle}, the encrypted frame is encrypted again via feedback $K_1$. The process is repeated $M$ times. After $M$-rounds of encryption, $K_2$ is turned on and this frame of the speech signal at the final round encryption is outputted by the channel. Then, $K_1$ is located at position 1 again, the next frame of the speech signal is encrypted similarly.
By this way, the $M$-rounds of encryption is done frame by frame.

(2) At the receiving end, the operating principle of the $M$ rounds of decryption for each frame of the speech data is illustrated as follows: First, $K_3$ is located at position~1, a frame of signal is decrypted. Then $K_3$ is located at position~2, this frame of signal is decrypted again by feedback $K_3$. The process is repeated $M$ times. After $M$ rounds of encryption, $K_4$ is turned on and this frame of the speech signal at the final round decryption is outputted. Then, $K_3$ is located at position~1 again, the next frame of the speech signal is decrypted similarly.

According to Fig.~\ref{fig:Principle}, a chaotic system (\ref{eq:example}) is used for the encryption, the basic iterative equation of chaotic map-based encryption system at the sending end is
\begin{equation}
\left\{
\begin{IEEEeqnarraybox}[\IEEEeqnarraystrutmode\IEEEeqnarraystrutsizeadd{2pt}{2pt}][c]{rCl}
x_1^{(d)}(k+1) & = & a_{11}^{(d)}x_1^{d}(k)+a_{12}^{(d)}x_2^{r}(k)+a_{13}^{(d)}x_3^{d}(k),\\
x_2^{(d)}(k+1) & = & a_{21}^{(d)}p(k)+a_{22}^{(d)}x_2^{d}(k)+a_{23}^{(d)}x_3^{d}(k),\\
x_3^{(d)}(k+1) & = & a_{31}^{(d)}p(k)+a_{32}^{(d)}x_2^{d}(k)+a_{33}^{(d)}x_3^{d}(k)\\
               &   & {\;\;}    +\varepsilon^{(d)}\sin(\sigma^{(d)}p(k)).
\end{IEEEeqnarraybox}
\right.
\label{eq:encryptOne}
\end{equation}
At the receiver, one gets the decryption system, given by
\begin{equation}
\left\{
\begin{IEEEeqnarraybox}[
\IEEEeqnarraystrutmode
\IEEEeqnarraystrutsizeadd{2pt}
{2pt}
][c]{rCl}
x_1^{(r)}(k+1) & = & a_{11}^{(r)}x_1^{r}(k)+a_{12}^{(r)}x_2^{r}(k)+a_{13}^{(r)}x_3^{r}(k),\\
x_2^{(r)}(k+1) & = & a_{21}^{(r)}p(k)+a_{22}^{(r)}x_2^{r}(k)+a_{23}^{(r)}x_3^{r}(k),\\
x_3^{(r)}(k+1) & = & a_{31}^{(r)}p(k)+a_{32}^{(r)}x_2^{r}(k)+a_{33}^{(r)}x_3^{r}(k)\\
               &  & {\;\;}    +\varepsilon^{(r)}\sin(\sigma^{(r)}p(k)),
\end{IEEEeqnarraybox}
\right.
\label{eq:decryptOne}
\end{equation}
where
$p(k) = \bmod (\left\lfloor {x_1^{(d)}(k)} \right\rfloor ,{2^8}) \oplus s(k)$,
$s(k)$ is the $k$-th element of the input speech signal, and
$\oplus$ denotes exclusive OR operation.

When parameters of the both sides match, one has $a_{i,j}^{(d)}=a_{i,j}^{(r)}=a_{i,j}$, $(1\leq i, j\leq 3)$,
$\epsilon^{(d)}=\epsilon^{(r)}=\epsilon$, and $\sigma^{(d)}=\sigma^{(r)}=\sigma$.
First, according to the second and third equations of system~(\ref{eq:encryptOne}) and (\ref{eq:decryptOne}), the iterative equation corresponding to the error signal is
\begin{equation*}
\begin{pmatrix}
\Delta x_2(k+1)\\
\Delta x_3(k+1)
\end{pmatrix}
=
\begin{pmatrix}
a_{22} & a_{22}\\
a_{32} & a_{33}
\end{pmatrix} \cdot
\begin{pmatrix}
\Delta x_2(k)\\
\Delta x_3(k)
\end{pmatrix}.
\end{equation*}
Then, one can further get
\begin{equation}
\label{eq:iteration}
\begin{pmatrix}
\Delta x_2(k)\\
\Delta x_3(k)
\end{pmatrix}
=
\begin{pmatrix}
a_{22} & a_{22}\\
a_{32} & a_{33}
\end{pmatrix}^k \cdot
\begin{pmatrix}
\Delta x_2(0)\\
\Delta x_3(0)
\end{pmatrix},
\end{equation}
where $\Delta x_2(0)$ and $\Delta x_3(0)$ are initial values, $\Delta x_2=x_2^{(r)}(k)-x_2^{(d)}(k)$, and $\Delta x_3=x_3^{(r)}(k)-x_3^{(d)}(k)$.
Taking norm on both sides of Eq.~(\ref{eq:iteration}), one can get
\begin{equation}
\left\|\begin{pmatrix}
\Delta x_2(k)\\
\Delta x_3(k)
\end{pmatrix}\right\|
\leq
\left\|\begin{pmatrix}
a_{22} & a_{22}\\
a_{32} & a_{33}
\end{pmatrix}\right\|^k \cdot
\left\|\begin{pmatrix}
\Delta x_2(0)\\
\Delta x_3(0)
\end{pmatrix}\right\|.
\end{equation}
As the characteristic roots corresponding to $ \begin{pmatrix}
a_{22} & a_{22}\\
a_{32} & a_{33}
\end{pmatrix}$ located in the unit circle, one can assure that its norm is less than one from the Gershgorin circle theorem.
So, one has
\begin{equation}
\lim_{k\rightarrow \infty} \|\Delta x_i(k) \|=\lim_{k\rightarrow \infty} \| x_i^{(r)}(k)- x_i^{(d)}(k) \|=0,
\label{eq:limits}
\end{equation}
where $i=2, 3$.

Now, we further consider the convergence of the error variable $\Delta x_1(k)$. According to the first equation of system~(\ref{eq:encryptOne}) and (\ref{eq:decryptOne}), one can get error iterative equation as
\begin{equation}
\Delta x_1(k+1) = a_{11} \Delta x_1(k) + a_{12} \Delta x_2(k)+ a_{13} \Delta x_3(k),
\label{eq:iterativequation}
\end{equation}
where $\Delta x_1(k)=x_1^{(r)}(k)-x_1^{(d)}(k)$.

When $k\rightarrow \infty$, one can get $\Delta x_2(k)\rightarrow 0$ and $\Delta x_3(k)\rightarrow 0$ from
Eq.~(\ref{eq:limits}). So, one has
\begin{equation}
\lim_{k\rightarrow \infty} \Delta x_1(k+1)=\lim_{k\rightarrow \infty} a_{11}\Delta x_1(k).
\end{equation}
As $|a_{11}|<1$, one has
\begin{equation}
\lim_{k\rightarrow \infty} \|\Delta x_1\| =\lim_{k\rightarrow \infty} \| x_1^{(r)}(k)-x_1^{(d)}(k)  \|=0.
\label{eq:finalerror}
\end{equation}

Now, one can see that Eq.~(\ref{eq:encryptOne}) and (\ref{eq:decryptOne}) can be synchronized when the parameters are matched.
In the receiving end, the original speech signal can be correctly decrypted as
$\hat{s}(k)=s(k)\oplus \bmod (\left\lfloor {x_1^{(d)}(k)} \right\rfloor ,{2^8}) \oplus \bmod (\left\lfloor {x_1^{(r)}(k)} \right\rfloor ,{2^8})
=s(k)$. In addition, with the matched parameters, the case for multiple rounds of encryption has the similar results. Note that Eq.~(\ref{eq:limits}) and (\ref{eq:finalerror}) can assure exponential asymptotic convergence, synchronization errors can be ignored just after a few iterations, and the synchronization of the chaotic system at the sending end and the receiving end can be achieved quickly, i.e., the speech signal can be decrypted exactly.

\section{Design of a multicast scheme for multiuser wireless communication and its Hardware implementation}
\label{sec:multicast}

\begin{figure}[!htbp]
\centering
\includegraphics[width=0.8\figwidth]{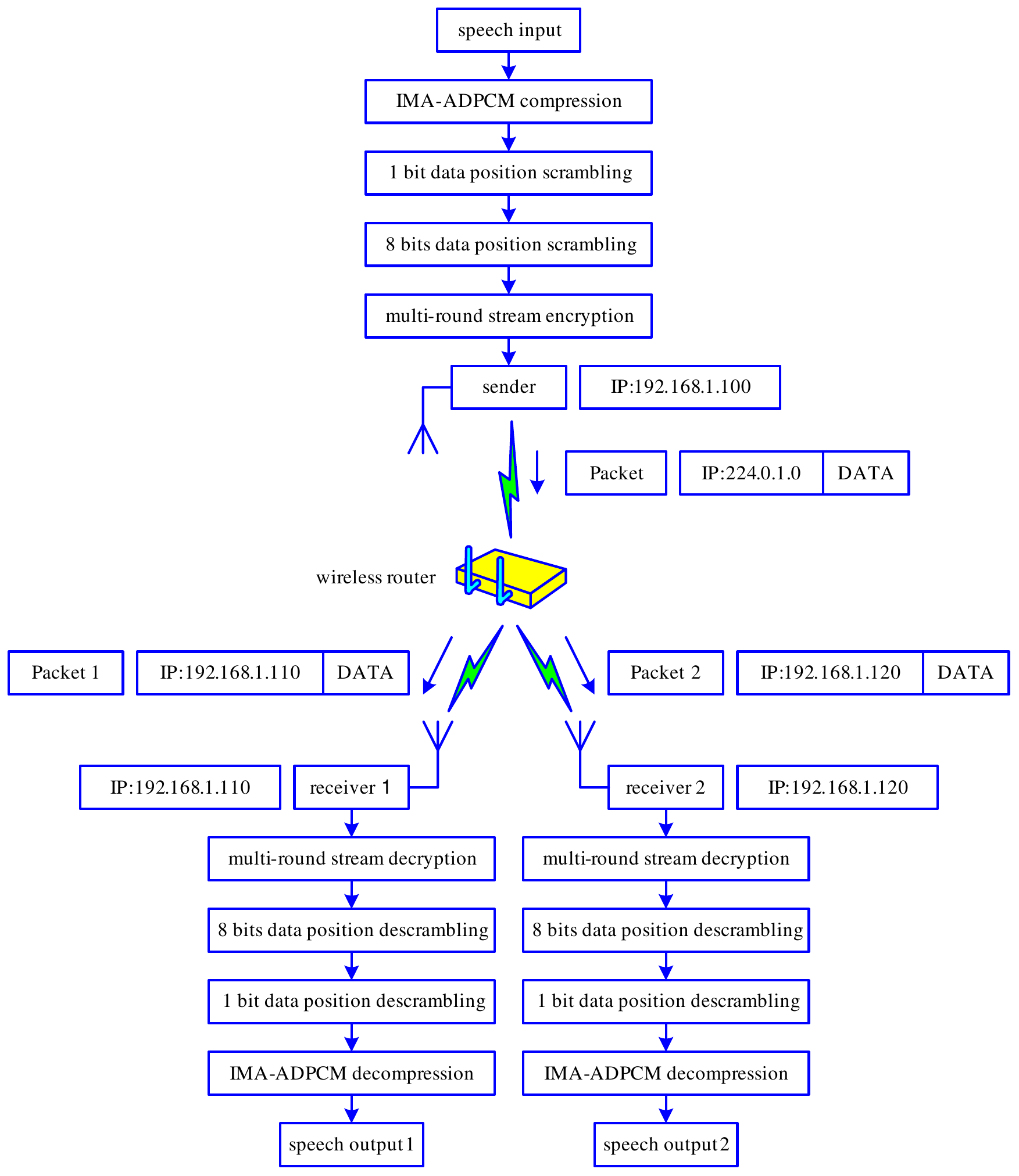}
\caption{The framework of the multicast-multiuser chaotic map-based speech communication system.}
\label{fig:framework}
\end{figure}

\begin{figure}[!htbp]
\center
\includegraphics[width=\figwidth]{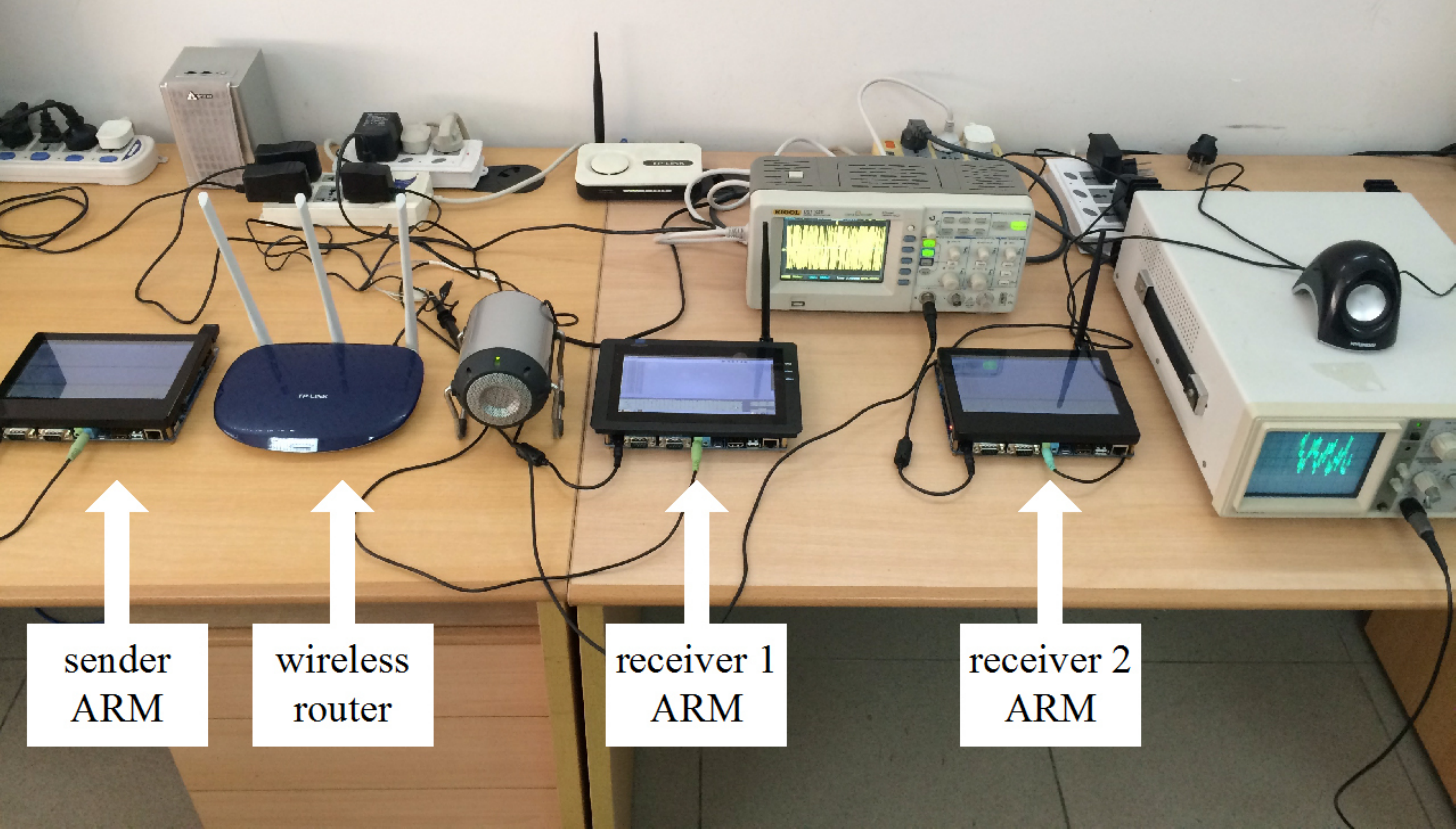}
\caption{The hardware implementation of the multicast-multiuser chaotic map-based speech communication system.}
\label{fig:hardware}
\end{figure}

The diagram of the multicast-multiuser chaotic map-based speech communication system is shown in Fig.~\ref{fig:framework}, and the corresponding hardware platform is shown in Fig.~\ref{fig:hardware}. At the sending end, after IMA-ADPCM compression of the input speech signal, three-level chaotic encryption scheme on speech communication is exerted. The specific encryption and its corresponding encryption keys are described as follows:

The first stage is that the position of every bit of the speech data is permuted randomly, the encryption key is the arrangement of the matrices given in Eq.~(\ref{eq:examplemat}), i.e.,
\begin{multline*}
\mathbf{T}_{12}\mathbf{T}_{13}\mathbf{T}_{14}\mathbf{T}_{15}\mathbf{T}_{16}\mathbf{T}_{23}\mathbf{T}_{24}\mathbf{T}_{25}\mathbf{T}_{26}
\mathbf{T}_{34}\mathbf{T}_{35}\mathbf{T}_{36}\mathbf{T}_{45}\mathbf{T}_{46}\mathbf{T}_{56}.
\end{multline*}
The second stage is scrambling of 8-bit data position, the encryption key is the arrangement of the matrix (\ref{eq:examplematn7}), i.e.,
\begin{multline*}
\mathbf{T}_{12}\mathbf{T}_{13}\mathbf{T}_{14}\mathbf{T}_{15}\mathbf{T}_{16}\mathbf{T}_{17}\mathbf{T}_{23}\mathbf{T}_{24}\mathbf{T}_{25}\mathbf{T}_{26}
\mathbf{T}_{27}\mathbf{T}_{34}\mathbf{T}_{35}\mathbf{T}_{36}\mathbf{T}_{37}\mathbf{T}_{45}\mathbf{T}_{46}\mathbf{T}_{47}\mathbf{T}_{56}\mathbf{T}_{57}
\mathbf{T}_{67}.
\end{multline*}
The third stage is encrypting speech data values by multi-round stream cipher, the encryption key of parameter matrix is
\[
\mathbf{A}=
\left(
  \begin{array}{ccc}
    0.205  & -0.595 & 0.265 \\
    -0.265 & -0.125 & 0.595 \\
    0.33   & -0.33  & 0.47
  \end{array}
\right).
\]

\begin{figure}[!t]
\center
\begin{minipage}{0.48\figwidth}
\center
\includegraphics[width=0.48\figwidth]{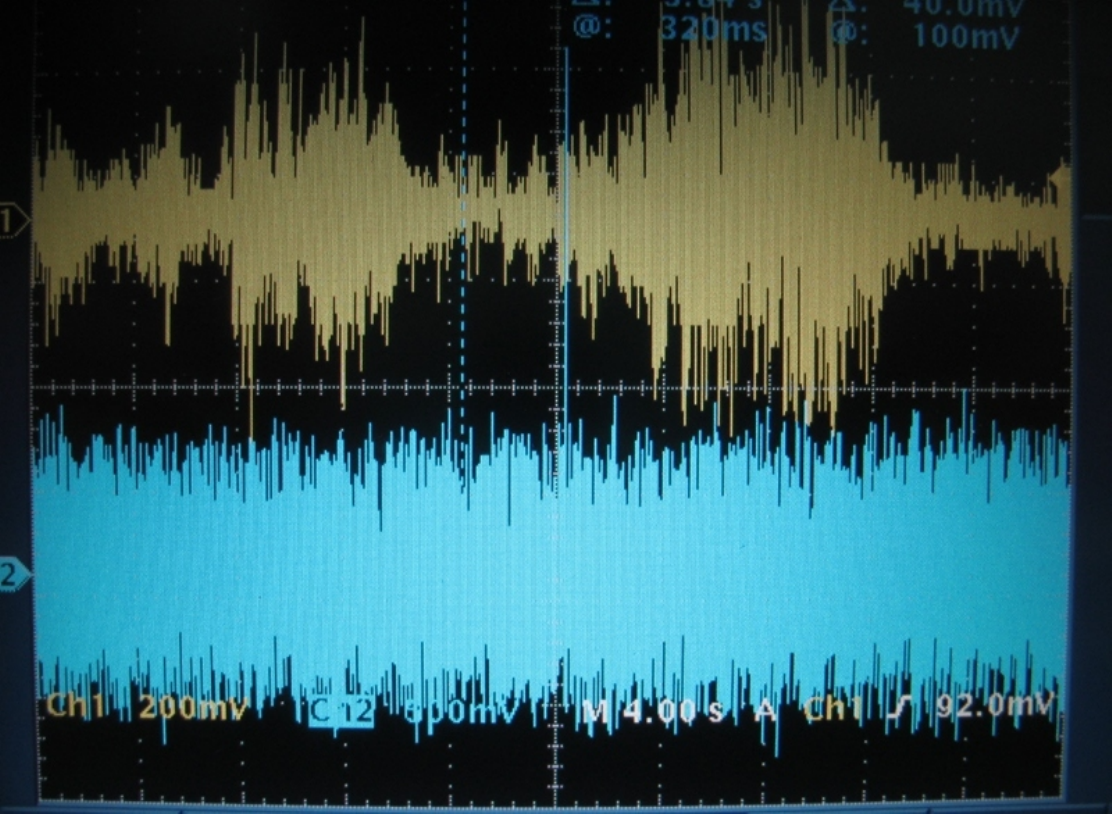}
a)
\end{minipage}
\begin{minipage}{0.48\figwidth}
\center
\includegraphics[width=0.48\figwidth]{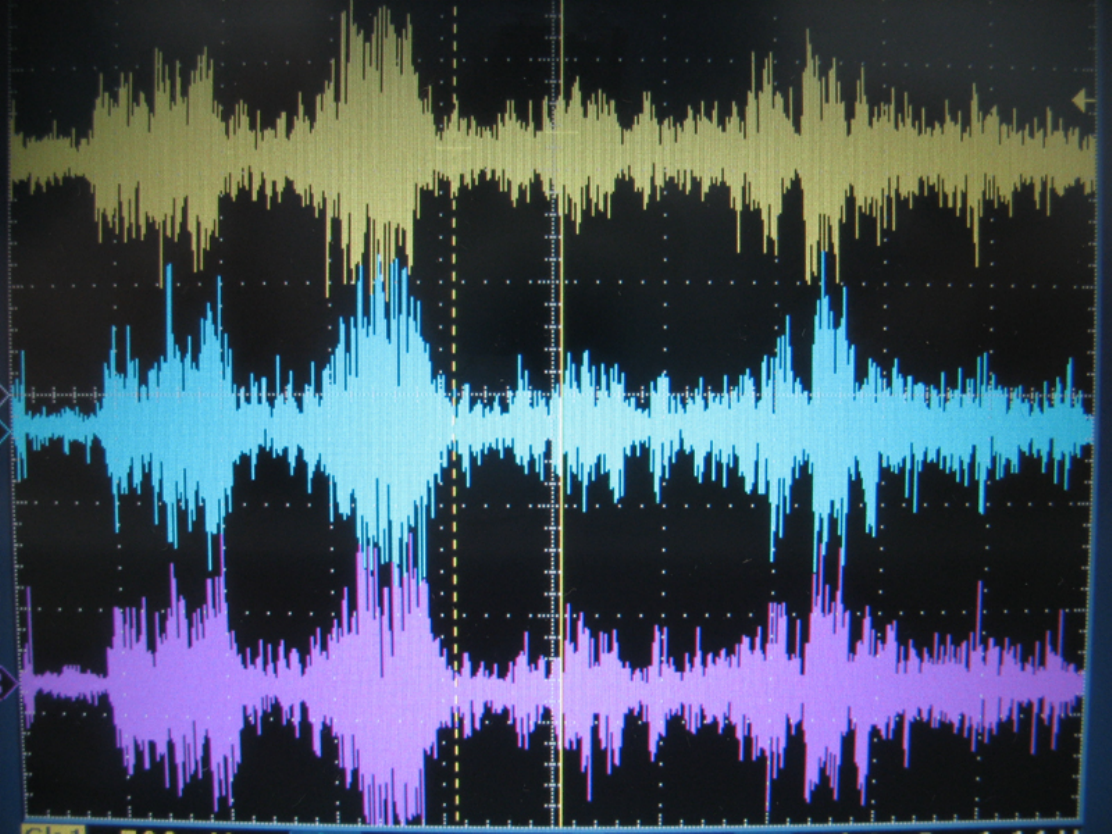}
b)
\end{minipage}\\
\begin{minipage}{0.48\figwidth}
\center
\includegraphics[width=0.48\figwidth]{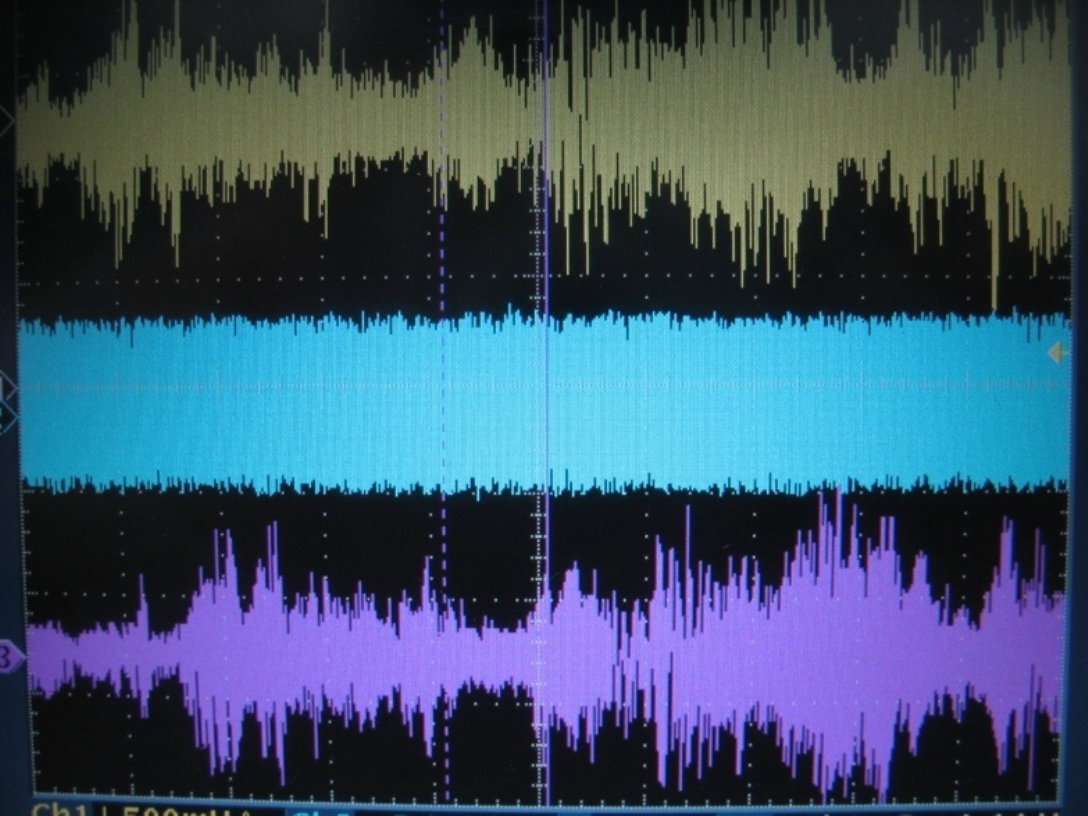}
c)
\end{minipage}
\begin{minipage}{0.48\figwidth}
\center
\includegraphics[width=0.48\figwidth]{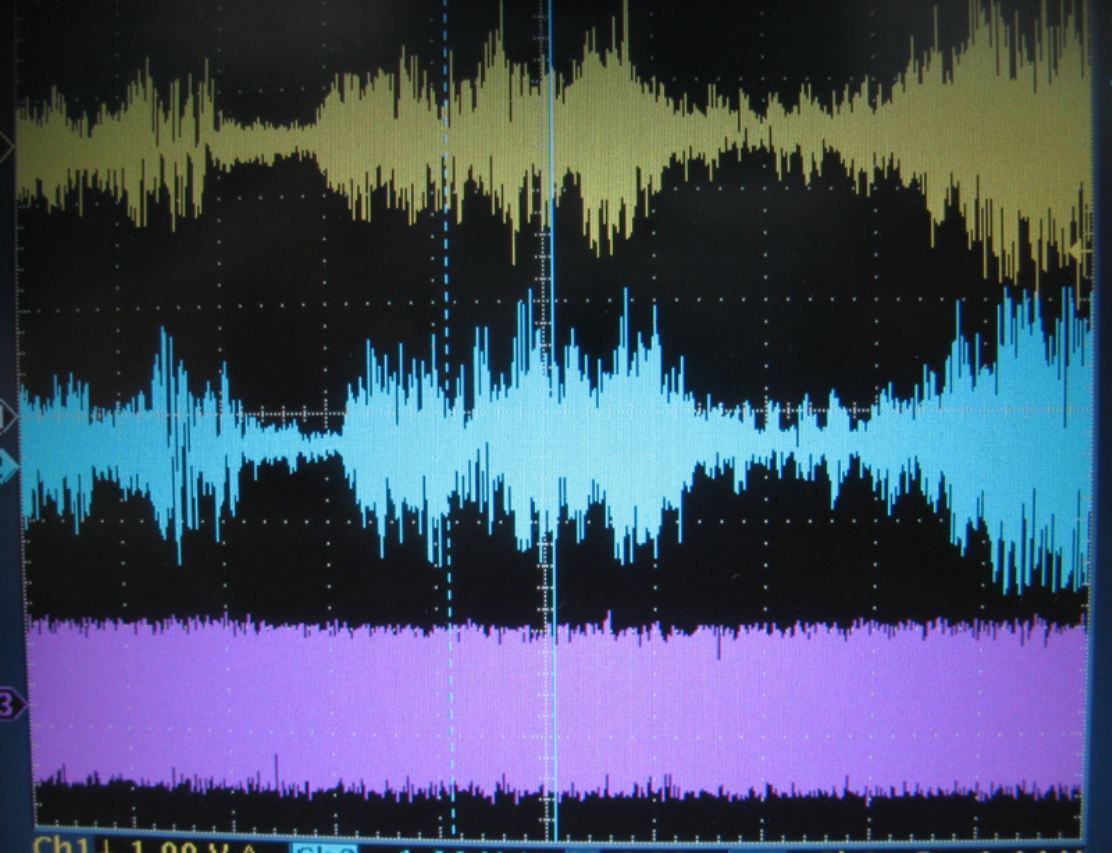}
d)
\end{minipage}
\caption{The original speech signal and the encrypted speech signal.}
\label{fig:results}
\end{figure}

At the receiving end, the original speech signal can be recovered by the corresponding inverse process only
when all the decryption key at the receiving end matches the encryption key at the sending end.
In this experiment, we select number of rounds for encryption $M=5$.
The construction of multicast-multiuser includes  one server at sending end and two clients at the receiving end (note that this construction of multicast-multiuser is also applicable for multiple clients at receiving end), the hardware experimental results are obtained as follows:

(1) The ARM implementation of the original speech signal input and the speech signal with three-level chaotic encryption after IMA-ADPCM compression are shown in the top and the bottom of Fig.~\ref{fig:results}a), respectively.

(2) When all the keys for the server at sending end and two clients at receiving end are matched, the ARM implementation results for the original speech signal, the decrypted speech signal at the first receiving end and the decrypted speech signal at the second receiving, are shown in the three rows
of Fig.~\ref{fig:results}b), respectively.

(3) All the keys for the server at sending end and the second client at receiving end are matched, but one decryption key for one round of decryption for the first client at the receiving end mismatches the encryption key at the receiving end. These lead to the following three conditions:
\begin{enumerate}
\item The first level encryption key for the server at sending end is
\begin{multline*}
\mathbf{T}_{12}\mathbf{T}_{13}\mathbf{T}_{14}\mathbf{T}_{15}\mathbf{T}_{16}\mathbf{T}_{23}\mathbf{T}_{24}\mathbf{T}_{25}\mathbf{T}_{26}
\mathbf{T}_{34}\mathbf{T}_{35}\mathbf{T}_{36}\mathbf{T}_{45}\mathbf{T}_{46}\mathbf{T}_{56}.
\end{multline*}
Then, the mismatched key of the first level decryption for the first client at receiving end is
\begin{multline*}
\mathbf{T}_{34}\mathbf{T}_{13}\mathbf{T}_{14}\mathbf{T}_{15}\mathbf{T}_{16}\mathbf{T}_{23}\mathbf{T}_{24}\mathbf{T}_{25}\mathbf{T}_{26}
\mathbf{T}_{12}\mathbf{T}_{35}\mathbf{T}_{36}\mathbf{T}_{45}\mathbf{T}_{46}\mathbf{T}_{56}.
\end{multline*}

\item The second level encryption key for the server at sending end is
\begin{multline*}
\mathbf{T}_{12}\mathbf{T}_{13}\mathbf{T}_{14}\mathbf{T}_{15}\mathbf{T}_{16}\mathbf{T}_{17}\mathbf{T}_{23}\mathbf{T}_{24}\mathbf{T}_{25}\mathbf{T}_{26}
\mathbf{T}_{27}\mathbf{T}_{34}\mathbf{T}_{35}\mathbf{T}_{36}\mathbf{T}_{37}\mathbf{T}_{45}\mathbf{T}_{46}\mathbf{T}_{47}\mathbf{T}_{56}\mathbf{T}_{57}
\mathbf{T}_{67}.
\end{multline*}
Then, the mismatched key for the second level decryption for the first client at receiving end is
\begin{multline*}
\mathbf{T}_{45}\mathbf{T}_{13}\mathbf{T}_{14}\mathbf{T}_{15}\mathbf{T}_{16}\mathbf{T}_{17}\mathbf{T}_{23}\mathbf{T}_{24}\mathbf{T}_{25}\mathbf{T}_{26}
\mathbf{T}_{27}\mathbf{T}_{34}\mathbf{T}_{35}\mathbf{T}_{36}\mathbf{T}_{37}\mathbf{T}_{12}\mathbf{T}_{46}\mathbf{T}_{47}\mathbf{T}_{56}\mathbf{T}_{57}
\mathbf{T}_{67}.
\end{multline*}

\item The key for the third level encryption for the server at sending end is
\[
\mathbf{A}=
\left(
  \begin{array}{ccc}
    0.205  & -0.595 & 0.265 \\
    -0.265 & -0.125 & 0.595 \\
    0.33   & -0.33  & 0.47
  \end{array}
\right).
\]
\end{enumerate}
Then, the mismatched key for the third level decryption for the first client at receiving end is
$\Delta a_{11}=10^{-9}$ instead of $a_{11}$. When there is any given mismatched key at any level of decryption, the ARM implementation results for the original speech signal, the decrypted speech signal at the first receiving end and the decrypted speech signal at the second receiving, are obtained in the three rows of Fig.~\ref{fig:results}c), respectively.

(4) All the keys for the server at sending end and the first client at receiving end are matched, but one decryption key for one round of decryption for the second client at the receiving end mismatches the encryption key at the receiving end. These cause the similar three conditions as above. Then, the ARM implementation results for the original speech signal, the decrypted speech signal at the first receiving end and the decrypted speech signal at the second receiving end, are obtained in the three rows of Fig.~\ref{fig:results}d), respectively. The video recordings of the ARM-embedded hardware experimental results can be accessed in the companion page of \cite{LinYuLuCaiChen15}.

\section{Analysis of security performance}
\label{sec:performance}

\subsection{Test results with NIST test suite}

In the test suite designed by National Institute of Standards and Technology in \cite{Rukhin:TestPRNG:NIST},
100 sequences of length $10^6$ bits, the $P$-value fall in the interval $[0.0001, 1]$, it is considered that the
random bit sequence generator pass the test, and fail otherwise.

In the encryption scheme composed of multi-round stream cipher shown in Fig.~\ref{fig:Principle}, after the iterative sequence is converted into 64-bit integers, its eight least significant bits of the speech signal are encrypted. Therefore, the $1\times 10^8$ bits sequence is made up of every eight least significant bits of iterative sequence in the corresponding NIST test suite, the test results are shown in Table~\ref{table:test}, where symbol ``-" indicates that the corresponding test failed. The relationship between the test results and the Lyapunov exponents is explained as follows:

(1) The results of P-value (1) with the corresponding parameter $\varepsilon=100$ and $\sigma=0.0667$, the Lyapunov exponents are obtained as $LE_1=-0.0017$,  $LE_2=-0.0993$, $LE_3=-0.2229$, respectively. Since the Lyapunov exponents are all negative, the generated sequence can not pass any item in the NIST test suite.

(2) The results of P-value (2) with the corresponding parameter $\varepsilon=300$ and $\sigma=0.2$, the Lyapunov exponents are obtained as $LE_1=1.1041$,  $LE_2=0.9588$, $LE_3=0.1828$, respectively. Since the Lyapunov exponents are not big enough, only two items in the NIST test suite can be passed.

(3) The results of P-value (3) with the corresponding parameter $\varepsilon=3000$ and $\sigma=2$, the Lyapunov exponents are obtained as $LE_1=3.3825$,  $LE_2=3.287$, $LE_3=0.1893$, respectively. Only nine items in the NIST test suite can be passed.

(4) The results of P-value (4) with the corresponding parameter $\varepsilon=30000$ and $\sigma=2$, the Lyapunov exponents are obtained as $LE_1=5.6879$,  $LE_2=5.65$, $LE_3=0.1895$, respectively. Only 13 items in the NIST test suite can be passed.

(5) The results of P-value (5) with the corresponding parameter $\varepsilon=3\cdot 10^5$ and $\sigma=200$, the Lyapunov exponents are obtained as $LE_1=7.9827$,  $LE_2=7.9814$, $LE_3=0.1895$, respectively. In this case, fourteen items in the NIST test suite can be passed.

(6) The results of P-value (6) with the corresponding parameter $\varepsilon=3\cdot 10^8$ and $\sigma=2\cdot 10^5$, the Lyapunov exponents are obtained as $LE_1=14.8403$,  $LE_2=14.8107$, $LE_3=0.1903$, respectively. Since the Lyapunov exponents are all positive and big enough, all items in the NIST test suite can be passed.

As shown in Table~\ref{table:test}, the value of positive Lyyapunov exponents
increases and degeneracy disappears as increase of $\epsilon$ and $\sigma$, so the generated sequence can pass all test items.

\setlength{\tabcolsep}{2pt}

\begin{table*}[!t]
\centering
\caption{The performed tests and the ratio of sequences passing each test in a sample of 100 sequences.}
\begin{tabular}{lllllll}
\hline\hline
Statistical Test          & P-value (1) & P-value (2) & P-value (3) & P-value (4) & P-value (5) & P-value (6) \\
Frequency                 & --           & --           & 0.946308    & 0.867692    & 0.798139    & 0.759756    \\
Block Frequency           & --           & --           & 0.035174    & 0.062821    & 0.574903    & 0.437274    \\
Cumulative Sums           & --           & --           & 0.7886635   & 0.448529    & 0.67685     & 0.2636575   \\
Runs                      & --           & --           & 0.574903    & 0.12962     & 0.035174    & 0.616305    \\
Long Runs of Ones         & --           & --           & 0.455937    & 0.249284    & 0.090936    & 0.534146    \\
Rank                      & --           & 0.108791    & 0.289667    & 0.946308    & 0.181557    & 0.946308    \\
FFT                       & --           & --           & 0.514124    & 0.534146    & 0.699313    & 0.096578    \\
Non-overlapping Templates & --           & --           & --           & --           & --           & 0.528973    \\
Overlapping Templates     & --           & --           & 0.249284    & 0.816537    & 0.924076    & 0.191687    \\
Universal                 & --           & --           & 0.867692    & 0.55442     & 0.181557    & 0.474986    \\
Approximate Entropy       & --           & --           & --           & 0.319084    & 0.595549    & 0.739918    \\
Random Excursions         & --           & --           & --           & 0.3943705   & 0.35054825  & 0.68632925  \\
Random Excursions Variant & --           & --           & --           & --           & 0.344756    & 0.6600216   \\
Serial                    & --           & --           & --           & 0.428899    & 0.0458625   & 0.7494675   \\
Linear Complexity         & --           & 0.366918     & --           & 0.991468    & 0.935716    & 0.032923    \\
Success                   & 0/15            & 2/15         &  9/15        & 13/15       & 14/15       & 15/15       \\
\hline \hline
\end{tabular}
\label{table:test}
\end{table*}

\subsection{Statistical analysis}

\begin{figure}[!htbp]
\center
\includegraphics[width=\figwidth]{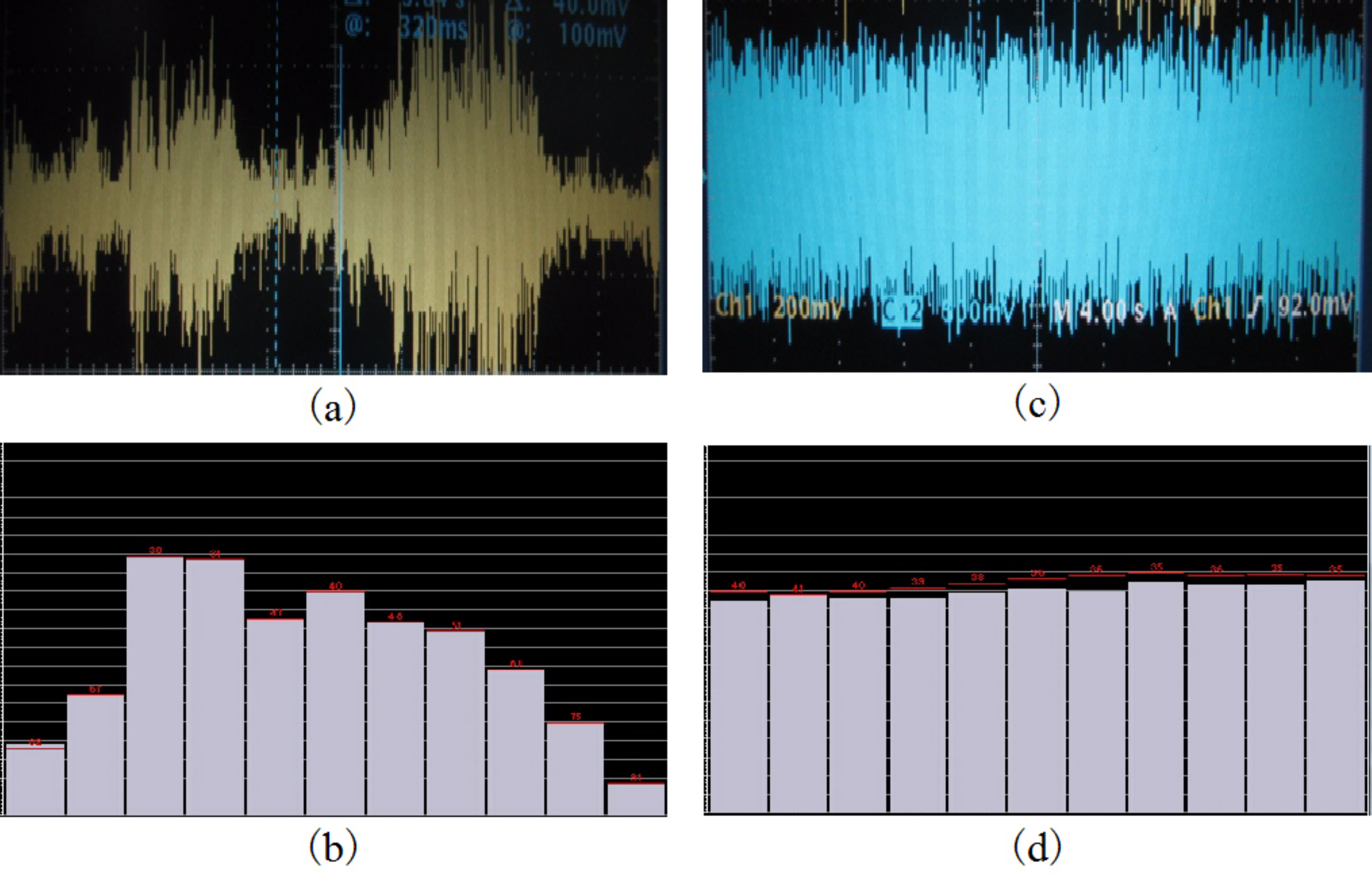}
\caption{The statistical analysis of one frame of speech data before and after the encryption by multi-round stream cipher: (a) The original speech signal; (b) The distribution of the original speech signal; (c) The speech signal encrypted by multi-round stream cipher; (d) The distribution of the speech signal encrypted by multi-round stream cipher.}
\label{fig:histo}
\end{figure}

Taking one frame of speech data for example, the statistical analysis of this frame of speech data before and after the encryption by multi-round stream cipher is performed to obtain the corresponding histogram shown in Fig.~\ref{fig:histo}.

As presented in Fig.~\ref{fig:histo}, the chaotic sequence owns good randomness, so the encrypted audio data approaches
uniform distribution. Thus, the proposed encryption method makes the ciphertext own weak correlation with the secrete key,
\[
H(K|C_1C_2\cdots, C_n)\approx H(K),
\]
where $H$ is the information entropy, $K$ is the secret key, $C_1C_2\cdots, C_n$ are cipher elements.
Thus, it increases the capability of withstanding statistical attacks and ciphertext-only attack much.

\subsection{Analysis of performance against differential attack}

The differential analysis plays an important role in the security analysis of chaotic cryptosystem, it can demonstrate whether the encryption results are sensitive to change of the plaintext, namely the ``avalanche effect". If the chaotic cryptosystem can withstand the differential attack, any variation in the plaintext could cause serious affect in the ciphertext. Such impact is widely measured by tow factors, NPCR (Number of Pixels Change Rate) and UACI (Unified Average Changing Intensity) \cite{YuLuChen13},
\begin{equation}
\left\{
\begin{IEEEeqnarraybox}[
\IEEEeqnarraystrutmode\IEEEeqnarraystrutsizeadd{2pt}{2pt}][c]{rCl}
\mbox{NPCR} & = & \frac{\sum\limits_{i}D(i)}{F}\cdot 100\%,\\
\mbox{UACI} & = & \frac{1}{F}\cdot \sum_{i}\frac{|C(i)-C'(i)|}{255}\cdot 100\%,
\end{IEEEeqnarraybox}
\right.
\label{eq:NpcrUaci}
\end{equation}
where
\begin{equation*}
\left\{
\begin{IEEEeqnarraybox}[
\IEEEeqnarraystrutmode\IEEEeqnarraystrutsizeadd{2pt}{2pt}][c]{rCl}
C  &=& [C_1 C_2 \cdots C_F]\\
C' &=& [C'_1 C'_2 \cdots C'_F],
\end{IEEEeqnarraybox}
\right.
\end{equation*}
\begin{equation*}
D(i)=
\begin{cases}
1   &  \text{if } C(i)\neq C'(i),\\
0   &  \text{if } C(i)= C'(i),
\end{cases}
\end{equation*}
$C(i)$ and $C'(i)$ denote the $i$-th element of two cipheretext $C$ and $C'$, respectively.

To study influence of each element of the original speech data for the corresponding ciphertext, a frame of speech data of 16384 elements is selected to calculate NPCR and UACI corresponding to each element, then the average value is calculated as
\begin{equation}
\left\{
\begin{IEEEeqnarraybox}[
\IEEEeqnarraystrutmode\IEEEeqnarraystrutsizeadd{2pt}{2pt}][c]{rCl}
\overline{\mbox{NPCR}} & = & \sum_{m=1}^{16384} \mbox{NPCR}(m)/16384,\\
\overline{\mbox{UACI}} & = & \sum_{m=1}^{16384} \mbox{UACI}(m)/16384,
\end{IEEEeqnarraybox}
\right.
\label{eq:averageNPCR}
\end{equation}
where $\mbox{NPCR}(m)$ and $\mbox{UACI}(m)$ denote the NPCR value and UACI value of the $m$-th element.

According to Eq.~(\ref{eq:averageNPCR}), it is shown in Fig.~\ref{fig:Principle} that $\overline{\mbox{NPCR}}=49.8\%$ and $\overline{\mbox{UACI}}=16.73\%$ after first round of encryption. When $M=2$, it is shown in Fig.~\ref{fig:Principle} that $\overline{\mbox{NPCR}}=99.61\%$ and $\overline{\mbox{UACI}}=33.47\%$ after second round encryption. In addition, $\overline{\mbox{NPCR}}$ and $\overline{\mbox{UACI}}$ become more and more larger with increase of round number.

By comparison of the above results, it is obvious that the ability of one-round encryption to resist differential attack is weaker than that of two-round encryption. The underlying mechanisms are detailed below.

In the method shown in Fig.~\ref{fig:Principle}, the sequence is encrypted from left to right in the encryption of one frame of speech data by one-round chaotic stream cipher, encryption for one speech frame is arranged as $s(1)\rightarrow s(2)\rightarrow s(3), \cdots, s(16383)\rightarrow s(16384)$. Then, after 16384 iterations, the one-round encryption for one frame of speech data is completed.

Figure~\ref{fig:Principle} shows the encryption and decryption process based on chaotic stream cipher, whose main feature is the use of a closed loop feedback system. That means the encrypted results should be fed back to the original one through the feedback system after each encryption, and makes the speech data and chaotic system ``all rolled into one". Thus, any slight variation in the plaintext data has a significant impact on the ciphertext.

However, there is a problem with the one-round encryption scheme: there are obvious differences among the impacts on the ciphertext caused by the variation in the different positions of the plaintext data. Specifically, the variation in data of the smaller index $i$ has stronger influence on the ciphertext, and the calculated NPCR and UACI are relatively big. The underlying reason is that the element of smaller index $i$ is encrypted earlier, the encrypted result can be earlier feedback to the original system through the feedback loop. More iterations the feedback to the original system are, more easier the data and the chaotic system combine. Thus the impacts on the ciphertext caused by the variation in the position are greater. The element of bigger index is encrypted later, and the encrypted results are feedback to the original system later. The more less the feedback to the original system iterates, the more harder efficient combination of the data and the chaotic system becomes. Thus, the impacts on the ciphertext caused by the variation in the position are smaller.

If multi-round encryption scheme is used, the results are different from one-round encryption. For example, in the case of more than two or two rounds encryption scheme, even for speech data $s(16384)$, the encrypted results also have the opportunity to mix efficiently with chaotic system in the next round of the encryption through the feedback loop to the original system. It is a fundamental solution for the problem that the impacts on the ciphertext caused by the variation in the bigger index are small.

\subsection{Sensitivity with respect to change of parameters of the discrete chaotic system}

Any secure encryption system is expected to own ``avalanche effect" with respect to change of secret key, i.e. tiny change of one sub-key can cause dramatic change of the corresponding ciphertext. The smaller magnitude of the mismatch errors, the better security of the system. In this subsection, we analyzed parameter sensitivity of the proposed system under different round numbers.

\subsubsection{The case under round number of one}

Set the absolute value of error caused by the mismatched key parameters between the receiving end and the sending end as
$|\Delta a_{i,j}|=|a_{i,j}^{(r)}-a_{i,j}^{(d)}|$, $1\leq i, j\leq 3$. Suppose that there are only one mismatched key parameter in
$|\Delta a_{i,j}|$ and others are matched. As the original speech signal cannot be decrypted, it was tested that the error matrix caused by one-round encryption scheme shown in Fig.~\ref{fig:Principle} with mismatched key parameter is
\begin{equation}
|\Delta a|_{C1} \triangleq
\begin{pmatrix}
    |\Delta a|_{11} & |\Delta a|_{12} & |\Delta a|_{13} \\
    |\Delta a|_{21} & |\Delta a|_{22} & |\Delta a|_{23} \\
    |\Delta a|_{31} & |\Delta a|_{32} & |\Delta a|_{33}
\end{pmatrix}
\varpropto
\begin{pmatrix}
    10^{-5} & 10^{-5} & 10^{-5} \\
    \underline{5}       & 10^{-5} & 10^{-5} \\
    \underline{5}       & 10^{-5} & 10^{-5}
\end{pmatrix}.
\label{eq:keyparameter}
\end{equation}

The underlined part of the above formula indicates that the corresponding key parameter is not sensitive to the mismatch errors, which is mainly caused
by that each encrypted speech data $p(k)=\bmod (\left\lfloor {x_1^{(d)}(k)} \right\rfloor ,{2^8})\oplus s(k)$ is fed back to these items by the feedback loop.

According to Eq.~(\ref{eq:keyparameter}), one-round encryption scheme shown in Fig.~\ref{fig:Principle} exists two invalid key parameters and only seven valid key parameters, the absolute value of the error for mismatch valid key parameters is of magnitude $10^{-5}$. The underlying reasons are analyzed as follows:

According to Eq.~(\ref{eq:keyparameter}), the mismatch error of $|\Delta a_{21}|$ and $|\Delta a_{31}|$ are of the same magnitude, the other mismatch errors   $\Delta a_{ij}$ ($i\neq 2, 3$; $j\neq 1$) are also in the another same magnitude. Therefore, only need to study $|\Delta a_{31}|$ and $|\Delta a_{32}|$, the similar results of other mismatched error can be obtained.

(1) Set $\varepsilon^{(d)}=\varepsilon^{(r)}=\varepsilon$, $\sigma^{(d)}=\sigma^{(r)}=\sigma$, assume that there is only one mismatch error
$|\Delta a_{31}=a_{31}^{(r)}-a_{31}^{(d)}|$ in $a_{31}$, and the other parameters are matched, according to Eq.~(\ref{eq:encryptOne}) and (\ref{eq:decryptOne}), the iterative equation of the error signal is given as
\begin{equation}
\left\{
\begin{IEEEeqnarraybox}[\IEEEeqnarraystrutmode\IEEEeqnarraystrutsizeadd{2pt}{2pt}][c]{rCl}
\Delta x_1(k+1) & = & a_{11}\Delta x_1(k)+a_{12}\Delta x_2(k)+a_{13}\Delta x_3(k),\\
\Delta x_2(k+1) & = & a_{22}\Delta x_2(k)+a_{23}\Delta x_3(k),\\
\Delta x_3(k+1) & = & \Delta a_{31}p(k)+a_{32}\Delta x_2(k)+a_{33}\Delta x_3(k).
\end{IEEEeqnarraybox}
\right.
\end{equation}
When the error signal reaches a steady state, $\Delta x_i(k+1)=\Delta x_i(k)$ is satisfied for $i=1, 2, 3$. The above equation becomes
\begin{equation}
\left\{
\begin{IEEEeqnarraybox}[\IEEEeqnarraystrutmode\IEEEeqnarraystrutsizeadd{2pt}{2pt}][c]{rCl}
\Delta x_1(k) & = & a_{11}\Delta x_1(k)+a_{12}\Delta x_2(k)+a_{13}\Delta x_3(k),\\
\Delta x_2(k) & = & a_{22}\Delta x_2(k)+a_{23}\Delta x_3(k),\\
\Delta x_3(k) & = & \Delta a_{31}p(k)+ a_{32}\Delta x_2(k)+a_{33}\Delta x_3(k).
\end{IEEEeqnarraybox}
\right.
\end{equation}
Solving the above equation, one can obtain
\begin{equation}
\left\{
\begin{IEEEeqnarraybox}[\IEEEeqnarraystrutmode\IEEEeqnarraystrutsizeadd{2pt}{2pt}][c]{rCl}
\Delta x_1(k) & = & -0.0887\cdot \Delta a_{31} p(k),\\
\Delta x_2(k) & = & 0.7507\cdot \Delta a_{31} p(k),\\
\Delta x_3(k) & = & 1.4194\cdot \Delta a_{31} p(k).
\end{IEEEeqnarraybox}
\right.
\label{eq:errorsoltion1}
\end{equation}

(2) Set $\varepsilon^{(d)}=\varepsilon^{(r)}=\varepsilon$, $\sigma^{(d)}=\sigma^{(r)}=\sigma$, assume that there is only one mismatch error
$|\Delta a_{32}=a_{32}^{(r)}-a_{32}^{(d)}|$ in $a_{32}$, and the other parameters are matched, $a_{ij}^{(r)}=a_{ij}^{(d)}=a_{ij}$ for $i\neq 3, j\neq 2$. According to Eq.~(\ref{eq:encryptOne}) and (\ref{eq:decryptOne}), the iterative equation with an error signal is given as
\begin{equation}
\left\{
\begin{IEEEeqnarraybox}[
\IEEEeqnarraystrutmode
\IEEEeqnarraystrutsizeadd{2pt}
{2pt}
][c]{rCl}
\overline{\Delta x_1}(k+1) & = & a_{11}\overline{\Delta x_1}(k)+a_{12}\overline{\Delta x_2}(k)+a_{13}\overline{\Delta x_3}(k),\\
\overline{\Delta x_2}(k+1) & = & a_{22}\overline{\Delta x_2}(k)+a_{23}\overline{\Delta x_3}(k),\\
\overline{\Delta x_3}(k+1) & = & a_{32}\overline{\Delta x_2}(k)+\Delta a_{32}  x_2(k)+ a_{33}\overline{\Delta x_3}(k).
\end{IEEEeqnarraybox}
\right.
\end{equation}
When the error signal reaches a steady state, $\overline{\Delta x_1}(k+1)=\overline{\Delta x_1}(k)$ is satisfied for $i=1, 2, 3$.
The above equation becomes
\begin{equation}
\left\{
\begin{IEEEeqnarraybox}[
\IEEEeqnarraystrutmode
\IEEEeqnarraystrutsizeadd{2pt}
{2pt}
][c]{rCl}
\overline{\Delta x_1}(k) & = & a_{11}\overline{\Delta x_1}(k)+a_{12}\overline{\Delta x_2}(k)+a_{13}\overline{\Delta x_3}(k),\\
\overline{\Delta x_2}(k) & = & a_{22}\overline{\Delta x_2}(k)+a_{23}\overline{\Delta x_3}(k),\\
\overline{\Delta x_3}(k) & = & a_{32}\overline{\Delta x_2}(k)+ \Delta a_{32} x_2(k) + a_{33}\overline{\Delta x_3}(k).
\end{IEEEeqnarraybox}
\right.
\end{equation}

Solving the above set of equations, one can obtain
\begin{equation}
\left\{
\begin{IEEEeqnarraybox}[\IEEEeqnarraystrutmode\IEEEeqnarraystrutsizeadd{2pt}{2pt}][c]{rCl}
\overline{\Delta x_1}(k) & = & -0.0887\cdot \Delta a_{32} x_2(k),\\
\overline{\Delta x_2}(k) & = & 0.7507\cdot \Delta a_{32} x_2(k),\\
\overline{\Delta x_3}(k) & = & 1.4194\cdot \Delta a_{32} x_2(k).
\end{IEEEeqnarraybox}
\right.
\label{errorsoltion}
\end{equation}

According to Fig.~\ref{fig:phase}, the maximum magnitude of $x_2(k)$ is $10^8$. From Fig.~\ref{fig:Principle}, one can see that the maximum magnitude of $p(k)$ is $2^8$. Dividing the left part and the right part of Eq.~(\ref{eq:errorsoltion1}) with that of Eq.~(\ref{errorsoltion}), respectively,
one can get
\[
\frac{|\Delta a_{31}|}{|\Delta a_{32}|} \varpropto \frac{|x_2(k)|_{max}}{|p(k)|_{max}} \varpropto 10^5.
\]
This explains insensitivity of parameter mismatch and error matrix given in Eq.~(\ref{eq:keyparameter}).

\subsubsection{The case under multiple rounds}

In multi-round of encryption, for example, set $M=2$ in Fig.~\ref{fig:Principle}, it was tested that the error matrix caused by two-round encryption in Fig.~\ref{fig:Principle} with mismatched key parameter is
\begin{equation}
|\Delta a|_{C2} \varpropto
\begin{pmatrix}
    10^{-8} & 10^{-8} & 10^{-8} \\
    10^{-2} & 10^{-8} & 10^{-8} \\
    10^{-2} & 10^{-8} & 10^{-8}
\end{pmatrix}.
\label{eq:errorsoltioni}
\end{equation}

According to the above equation, two-round encryption in Fig.~\ref{fig:Principle} exists 9 key parameters, the errors of mismatched key parameters are of magnitude between $10^{-8}$ and $10^{-2}$. Similarly, set $M=5$ in Fig.~\ref{fig:Principle}, it was tested that the error matrix caused by five-rounds encryption in Fig.~\ref{fig:Principle} with mismatched key parameter is
\begin{equation}
|\Delta a|_{C5} \varpropto
\begin{pmatrix}
    10^{-9} & 10^{-9} & 10^{-9} \\
    10^{-3} & 10^{-9} & 10^{-9} \\
    10^{-3} & 10^{-9} & 10^{-9}
\end{pmatrix}.
\label{eq:errorsoltion2}
\end{equation}
In particular, the absolute value of error caused by the mismatched key parameters in Eq.~(\ref{eq:keyparameter}), (\ref{eq:errorsoltioni}), (\ref{eq:errorsoltion2}) does not need to be very accurate, and the coarse scope is enough. In addition, Eq.~(\ref{eq:keyparameter}) and (\ref{eq:errorsoltioni}), (\ref{eq:errorsoltion2}) mean that receiving end can't decrypt the original speech signal if error of any key parameter is greater than or equal to a given magnitude. In conclusion, by comparing Eq.~(\ref{eq:keyparameter}), (\ref{eq:errorsoltioni}), (\ref{eq:errorsoltion2}), we can conclude that the avalanche effect in the mismatch of key parameter is significantly enhanced with increase of the round number.

When the round number of encryption is increased, the key parameter becomes increasingly sensitive due to the extreme sensitivity of the chaotic system, and with increase of the number and quantity of positive Lyapunov exponents. In Fig.~\ref{fig:Principle}, at the receiving end, as long as there is a mismatched parameter, every parameter error after each round of encryption leads to error diffusion. But the proposed hyper chaotic system does not have degeneration, and the positive Lyapunov exponents are sufficiently large, thus the sensitivity of this parameter error is greater, and error diffusion is stronger.

\subsection{Estimation of the space of secret key}

According to the order of multiplication in each transformation sub-matrix $\mathbf{T}_{ij}$ of Eq.~(\ref{eq:matrixmultiply}), one has
\begin{equation}
\mathbf{A}=
\prod\limits_{i=1}^{n-1} {\left( {\prod\limits_{j = i + 1}^{n} {\mathbf{T}_{\alpha_i \beta_j}} } \right)}
\label{eq:matrixmultiply2}
\end{equation}
by ranking the subscript of $\mathbf{T}_{ij}$ with all possible orders, where $\alpha_i$ and $\beta_j$ should satisfy the following conditions:
\begin{equation}
\left\{
\begin{IEEEeqnarraybox}[\IEEEeqnarraystrutmode\IEEEeqnarraystrutsizeadd{2pt}{2pt}][c]{rcl}
\alpha_i & \in  & \{1, 2, \cdots, n-1\},\\
\alpha_p & \neq & \alpha_q \text{ if } p\neq q,\\
\beta_j  & \in  & \{2, 3, \cdots, n\} \text{ for } j=i+1, i+2, \cdots, n,\\
\beta_r  & \neq & \beta_s \text{ if } r\neq s,\\
\beta_j  & >    & \alpha_i.
\end{IEEEeqnarraybox}
\right.
\end{equation}

According to the above rules, the subscript of $\mathbf{T}_{\alpha_i \beta_j}$ in (\ref{eq:matrixmultiply2}) is sorted
as
\[
S=\alpha_1\beta_2,\alpha_1\beta_3,\cdots, \alpha_1\beta_n, \alpha_2\beta_3, \alpha_2\beta_4, \cdots, \alpha_2\beta_{n}, \cdots, \alpha_{n-1}\beta_n.
\]
In each exchange of the subscript $\alpha_i\beta_j$ of $\mathbf{T}_{\alpha_i \beta_j}$, we get a corresponding new subscript order, each of which can be used as a key for the speech encryption. According to the above ordering rule, the number of all possible orders is
$K_s^{(1)}={n \choose 2}$. Then according to order of priority $E(S, \alpha_i)$ in Eq.~(\ref{eq:numberPerS}), the total corresponding sort is
$K_s^{(2)}=n!$.

(1) In the first level of encryption, position of every bit of the compressed data is permuted randomly, set $n=6$, the number of possible order is
\[
K_s^1=K_s^{(1)}\cdot K_s^{(2)}= {6 \choose 2} \cdot 6!=9.4154\cdot 10^{14}.
\]

(2) In the second level of encryption, the obtained data is further permuted in the level of byte, set $n=7$, the number of all possible order is
\[
K_s^2=K_s^{(1)}\cdot K_s^{(2)}= {7 \choose 2} \cdot 7!=2.575\cdot 10^{23}.
\]

(3) In the third level of encryption, it is operated with a multi-round chaotic stream cipher. At 5-rounds encryption, the number of the valid key parameters is 9. When the error for mismatch valid key parameters $\{\Delta a_{i, j}\}_{i=1, j=1}^{3,3}$ is less than the value in the error matrix mismatch valid key parameters in Eq.~(\ref{eq:errorsoltion2}). The original speech signal can be decrypted. Therefore, the size of key space is
\[
K_s^3=\prod_{i=1}^3\left(\prod_{j=1}^3\frac{|a_{i, j}|}{|\Delta a_{i, j}|}\right)=3.2607\cdot 10^{64}.
\]

To sum up, the size of the key space for three-level encryption system is
\begin{equation*}
K_s=K_s^1\cdot K_s^2\cdot K_s^3=7.9055\cdot 10^{102}.
\end{equation*}

\section{Conclusion}

In this paper, a chaotic map-based multicast scheme for multiuser wireless communication was
designed and implemented in an ARM-embedded platform. Detailed hardware experiment results are
provided to show its real performance. Comprehensive theoretical analysis results support security
performance of the proposed secure speech wireless communication scheme.

\section*{Acknowledgement}

This work was supported by the National Key Research and Development Program of China
(No. 2016YFB0800401), the National Natural Science Foundation of China (No. 61671161,
61532020, 11072254), Science and Technology Planning Project of Guangzhou (No. 20151001036),
and Hunan Provincial Natural Science Foundation of China (no. 2015JJ1013).



\end{document}